\documentclass[3p, authoryear]{elsarticle}
\usepackage[utf8]{inputenc}
\usepackage{amsmath,amssymb,amsthm,bbold,bm}
\usepackage{graphicx}
\usepackage[font=small,labelfont=sf,textfont=sf]{subfig}
\usepackage[font=small,textfont=sf,labelfont={it,bf},margin=0cm,figurewithin=section,tablewithin=section]{caption}
\usepackage{color}
\definecolor{darkblue}{RGB}{16,0,160}
\definecolor{darkred}{RGB}{220,0,16}
\usepackage[colorlinks=true, citecolor=darkblue, pdfstartview=FitV, linkcolor=darkred]{hyperref}
\usepackage{physics}
\usepackage{enumerate}
\usepackage{calc}
\newsavebox\CBox
\newcommand\hcancel[2][0.5pt]{%
  \ifmmode\sbox\CBox{$#2$}\else\sbox\CBox{#2}\fi%
  \makebox[0pt][l]{\usebox\CBox}%
  \rule[0.5\ht\CBox-#1/2]{\wd\CBox}{#1}}

\usepackage{enumerate}
\usepackage{natbib}
\usepackage{url}
\usepackage[detect-weight=true, detect-family=true, detect-shape=true]{siunitx}
\usepackage{multirow,booktabs}
\usepackage[ruled,vlined]{algorithm2e}
\makeatletter
\g@addto@macro\bfseries{\boldmath}
\makeatother
\definecolor{newstuff}{RGB}{0,0,0}
\newcommand{\new}[1]{\textcolor{newstuff}{#1}}
\definecolor{newstuff_2nd_revision}{RGB}{0,0,0}
\newcommand{\newb}[1]{\textcolor{newstuff_2nd_revision}{#1}}

\numberwithin{equation}{section}

\theoremstyle{plain}
\newtheorem{theorem}{Theorem}[section]
\newtheorem{lemma}[theorem]{Lemma}

\theoremstyle{definition}

\newcommand{\bitem}{\begin{itemize}}
\newcommand{\eitem}{\end{itemize}}
\newcommand{\bal}{\begin{aligned}}
\newcommand{\eal}{\end{aligned}}
\newcommand{\beq}{\begin{equation}}
\newcommand{\eeq}{\end{equation}}
\newcommand{\mc}[1]{\mathcal{#1}}

\newcommand{\R}{\mathbb{R}}

\newcommand{\bpm}{\begin{pmatrix}}
\newcommand{\epm}{\end{pmatrix}}
\newcommand{\bsm}{\left(\begin{smallmatrix}}
\newcommand{\esm}{\end{smallmatrix}\right)}

\newcommand{\la}{\langle}
\newcommand{\ra}{\rangle}

\newcommand{\veps}{\varepsilon}

\newcommand{\vphi}{\varphi}

\DeclareMathOperator{\diag}{diag}

\DeclareMathOperator{\epi}{epi}

\DeclareMathOperator*{\argmax}{arg\,max\,}

\DeclareMathOperator{\conv}{conv}

\DeclareMathOperator{\logexp}{logexp}

\newcommand*\xbar[1]{%
  \hbox{%
    \vbox{%
      \hrule height 0.5pt 
      \kern0.3ex
      \hbox{%
        \kern-0.1em
        \ensuremath{#1}%
        \kern+0.1em
      }%
    }%
  }%
}

\title{Fast multivariate log-concave density estimation}
\author[1]{Fabian Rathke\corref{cor1}}
\ead{fabian.rathke@iwr.uni-heidelberg.de}
\author[1]{Christoph Schn\"{o}rr}
\ead{schnoerr@math.uni-heidelberg.de}
\cortext[cor1]{Corresponding author (Mail: fabian.rathke@iwr.uni-heidelberg.de/frathke@gmail.com, Tel.: +49 6221 5414858)}
\address[1]{$^2$Image \& Pattern Analysis Group (IPA), Heidelberg University, Im Neuenheimer Feld 205, 69120 Heidelberg, Germany.}

\begin{document}
\begin{abstract} 

A novel computational approach to log-concave density estimation is proposed. Previous approaches 
utilize the piecewise-affine parametrization of the density induced by the given sample set. The number of parameters as well as non-smooth subgradient-based convex optimization for determining the maximum likelihood density estimate cause long runtimes for dimensions $d \geq 2$ and large sample sets.
The presented approach is based on mildly non-convex smooth approximations of the objective function and \textit{sparse}, adaptive piecewise-affine density parametrization. Established memory-efficient numerical optimization techniques enable to process larger data sets for dimensions $d \geq 2$. While there is no guarantee that the algorithm returns the maximum likelihood estimate for every problem instance, we provide comprehensive numerical evidence that it does yield near-optimal results after significantly shorter runtimes. For example, 10000 samples in $\R^2$ are processed in two seconds, rather than in $\approx 14$~hours required by the previous approach to terminate. For higher dimensions, density estimation becomes tractable as well: Processing $10000$ samples in $\R^6$ requires 35 minutes. 
The software is publicly available as CRAN R package \texttt{fmlogcondens}. 
\end{abstract}
\maketitle

\section{Introduction}
\label{sec:introduction}
\subsection{Motivation, Related Work}
Log-concave density estimation has been an active area of research. Quoting \cite{chen2013}, the ``allure is the prospect of obtaining fully automatic nonparametric estimators, with no tuning parameters to choose'', as a flexible alternative to parametric models, like the Gaussian, that are often adopted by practitioners in an ad-hoc way. The mathematical analysis as well as the design of algorithms benefit from the convexity properties of the class of log-concave densities. We refer to \cite{Samworth:2017aa} for a recent survey.

\begin{figure}
\centering 
{\includegraphics[width=0.55\textwidth]{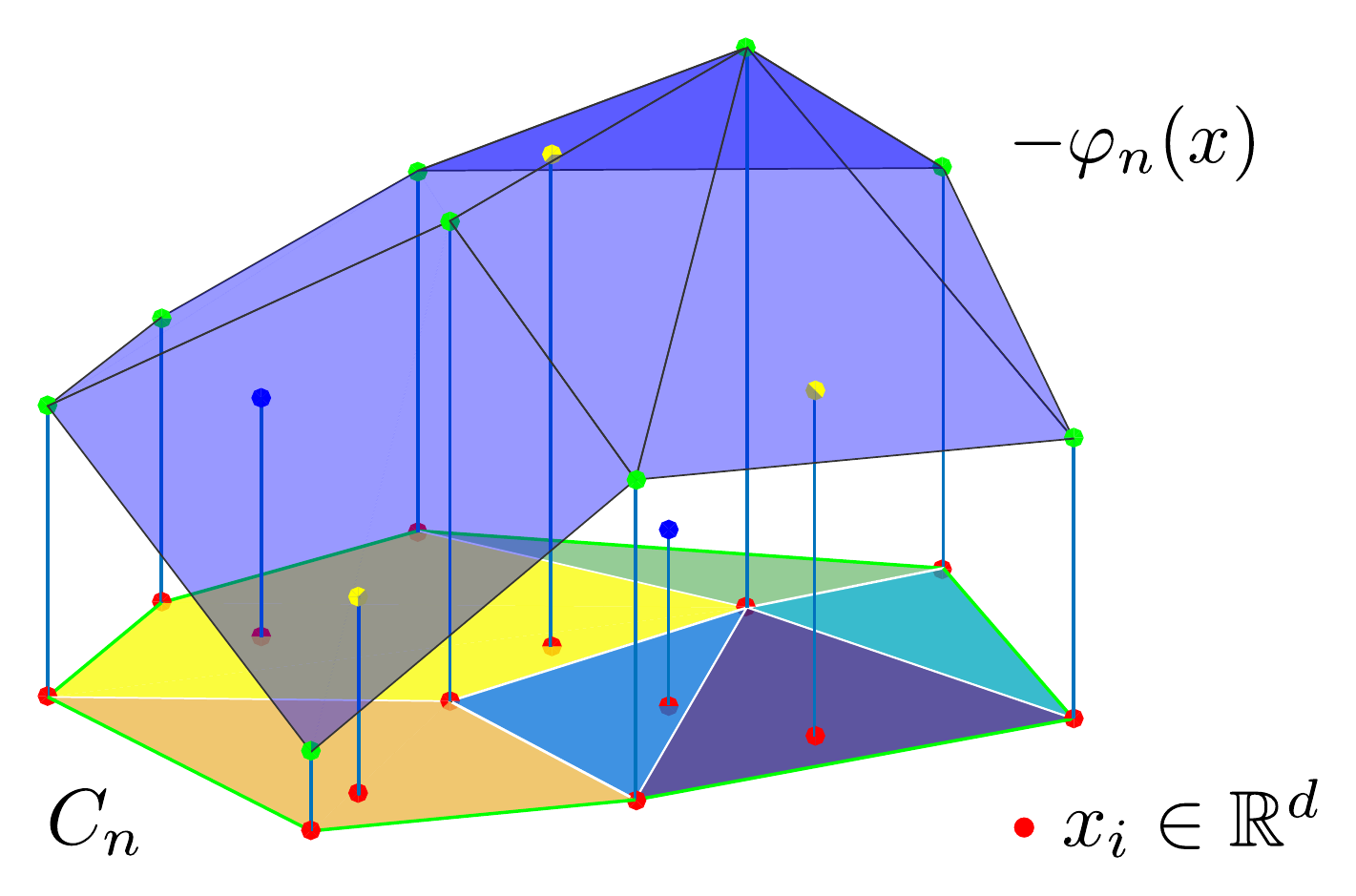}}
\caption{
A piecewise affine concave function $-\vphi_{n}(x)$ \eqref{eq:vphi-affine} whose parameters determine the density estimate \eqref{eq:def-hat-f}. The function values $-\vphi(x_{1}),\dotsc,-\vphi(x_{n})$ at given data points $\mc{X}_{n}=\{x_{1},\dotsc,x_{n}\} \subset \R^{d}$ induce a polyhedral decomposition $C_{n}$ of the convex hull $\conv(X)$, here shown for the case of bivariate data. Determining the parameters through iterative numerical optimization may change this decomposition depending on which data point defines a vertex of the hypograph (green point) or not (yellow and blue points). This increases the complexity considerably, in particular for larger $n$ and dimension $d$. In this paper, we work with a smooth approximation that enables more efficient log-concave density estimation.
}
\label{fig:tent-function}
\end{figure}

The general form of a log-concave density reads
\begin{equation}
f(x) = \exp\big(-\vphi(x)\big),\qquad
\vphi \in \mc{F}_{0}(\R^{d}),
\end{equation}
where $\mc{F}_{0}(\R^{d})$ denotes the class of convex lower-semicontinuous proper functions $\vphi \colon \R^{d} \to (-\infty,\infty]$ such that $\int_{\R^{d}}f = 1$. Given  i.i.d.~samples 
\begin{equation}
\mc{X}_{n} = \{x_{1},\dotsc,x_{n}\} 
\subset \R^{d}
\end{equation}
of a random vector $X \sim f$, with $n \geq d+1$, the task is to determine an estimate  
\begin{equation}\label{eq:def-hat-f}
\hat f_{n} = \exp\big(-\hat \vphi_{n}(x)\big)
\end{equation}
of $f$. This estimate,  
determined as maximizer of the log-likelihood, exists and is unique with probability $1$ \citep[Thm.~1]{cule2010}. Moreover, the corresponding convex function $\hat \vphi_{n} \in \mc{F}_{0}$ is supported on the convex hull $C_{n} = \conv \mc{X}_{n}$ of the given data and is \textit{piecewise linear}, in the sense of \citet[Def.~2.47]{rockafellar2009}: $C_{n}$ can be represented as union of finitely many polyhedral sets
\begin{equation}\label{eq:Cn-decomposition}
C_{n} = \bigcup_{i=1}^{N_{n,d}} C_{n,i},
\end{equation}
relative to each of which $\hat \vphi_{n}$ admits the \textit{affine} representation
\begin{equation}\label{eq:vphi-affine}
\hat\vphi_{n}(x)\big|_{C_{n,i}} =: \hat\vphi_{i,n}(x) = \la a_{i}, x \ra + b_{i},\qquad 
a_{i} \in \R^{d},\; b_{i} \in \R,\qquad
i =1,\dotsc,N_{n,d}.
\end{equation}
This is equivalent to the fact that the epigraph of $\hat \vphi_{n}$, denoted by 
\begin{equation} \label{eq:vphi-epi}
\epi \hat \vphi_{n} = \big\{(x,\alpha) \in \R^{d} \times \R \colon \alpha \geq \hat \vphi_{n}(x)\big\}
\end{equation}
is polyhedral and $\hat \vphi_{n}$ admits the representation \cite[Thm.~2.49]{rockafellar2009} 
\begin{equation}\label{eq:vphi-max}
\hat \vphi_{n} = \begin{cases}
\max\big\{\hat \vphi_{1,n}(x),\dotsc,\hat \vphi_{N_{n,d},n}(x)\big\},
& x \in C_{n}, \\
\infty, & x \not\in C_{n}.
\end{cases}
\end{equation}
We denote the class of piecewise linear proper convex functions over $C_{n}$ by 
\begin{equation}\label{eq:def-Phi}
\Phi_{n} := \big\{\vphi_{n} \in \mc{F}_{0}(\R^{d}) \colon \vphi_{n}\;\text{has the form \eqref{eq:vphi-affine} and \eqref{eq:vphi-max}}\big\}.
\end{equation}
Figure \ref{fig:tent-function} displays a function $\vphi_{n}$ in the planar case $d=2$. Given the function values
\begin{equation}
y_{\vphi} = (y_{\vphi,1},\dotsc,y_{\vphi,n}) := \big(\vphi_{n}(x_{1}),\dotsc,\vphi_{n}(x_{n})\big),
\end{equation}
$\vphi_{n}$ is uniquely determined as \textit{lower convex envelope}, that is the largest convex function majorized at the  given sample points $x_{i}$ by $y_{\vphi,i}$,
\begin{equation}
\vphi_{n}(x_{i}) \leq y_{\vphi,i},\qquad i=1,\dotsc,n.
\end{equation}

Due to \citet[Thm.~2]{cule2010}, a natural and admissible variational approach for determining the maximum likelihood estimate $\hat f_{n}$ in terms of $\hat \vphi_{n}$ and $\hat y_{\vphi}$, respectively, is given by
\begin{equation}\label{eq:def-J}
\hat y_{\vphi} = \arg\min_{y_{\vphi}} J(y_{\vphi}),\qquad
J(y_{\vphi}) = \frac{1}{n} \sum_{i=1}^{n} y_{\vphi,i}
+ \int_{C_{n}} \exp\big(-\vphi_{n}(x)\big)\dd{x}
\end{equation}
where the latter integral acts like a Lagrangian multiplier term enforcing the constraint $\int_{C_{n}} f_{n} = 1$ \cite[Thm.~3.1]{silverman1982}. In fact, it was shown that solving problem \eqref{eq:def-J} amounts to effectively minimizing over $\Phi_{n}$ \eqref{eq:def-Phi} to obtain $\hat\vphi_{n}$ and in turn the ML-estimate \eqref{eq:def-hat-f}. 

An algorithm for computing $\hat y_{\vphi}$ was worked out by \cite{cule2010} based on the convexity of $J$. While this algorithm is guaranteed to return the global optimum, its runtime complexity suffers from two facts:
\begin{enumerate}[(i)]
\item 
The objective function $J$ is convex but \textit{non-smooth} due to the polyhedral class of functions \eqref{eq:def-Phi} in which the algorithm searches for $\hat\vphi_{n}$. As consequence, the iterative scheme is based on subgradients which are known to converge rather slowly.
\item
The integral of \eqref{eq:def-J} has to be evaluated in every iterative step for each subset $C_{n,i}$ of the polyhedral decomposition \eqref{eq:Cn-decomposition}, where the subsets $C_{n,i}$ are required to be \textit{simplices}. While this can be conveniently done in closed form \cite[App.~B]{cule2010}, it is the increasing number of these subsets for larger dimension $d >2$ that slows down the algorithm.
\end{enumerate}
The number $N_{n,d}$ of components of the decomposition \eqref{eq:Cn-decomposition} is known to depend linearly on $n$, $N_{n,d} = \mc{O}(n)$, for $n$ points \textit{uniformly} distributed in $\R^{d}$ \citep{Dwyer:1991aa}, whereas the worst case bound for `pathologically' distributed $n$ points is $N_{n,d} = \mc{O}(n^{\lceil\frac{d}{2}\rceil})$, i.e.~grows exponentially with the dimension $d$ \citep{McMullen:1970aa}. For $n$ points sampled from log-concave distributions that are unimodal and in this sense simply shaped, it is plausible to assume that the lower complexity bound holds approximately, i.e.~a \textit{linear dependency} $N_{n,d} = \mc{O}(n)$. This means, in particular, that the number of parameters of the affine functions forming $\hat\vphi_{n}$ due to \eqref{eq:vphi-max} linearly depends on $n$ as well. 
While these bounds take into account the entire data set $\mc{X}_n$, it was shown for $d=1$ that under sufficient smoothness and other conditions, not all $x_i$ need to participate in the decomposition $C_n$ \citep{duembgen2009}. No proofs exist for $d>1$, but results presented in this paper indicate that this property of the ML estimator carries over to the multivariate case. Therefore the actual dependency of $N_{n,d}$ on $n$ may be lower than $\mc{O}(n)$.

On the other hand, concerning the ultimate objective of accurately estimating a multivariate log-concave density $f$, it was recently shown by \cite{Diakonikolas:2017aa} that in order to achieve an estimation error~$\epsilon$ in total variation distance with high probability, a function~$\hat{f}_n$ suffices that is defined by $\mc{O}\big((1/\epsilon)^{(d+1)/2}\big)$ hyperplanes. In the univariate case $d=1$, an algorithm that matches this complexity bound was published recently   
\citep{Acharya:2017aa}. In the multivariate case~$d>1$, on the other hand, the design of a computationally efficient algorithm was considered as a ``challenging and important open question'' by \cite{Diakonikolas:2017aa}.

\new{Quite recently, two approaches where published \citep{axelrod2018, diakonikolas2018} which solve the log-concave MLE \eqref{eq:def-J} with high probability with an estimation error $\epsilon < 1$ in terms of the total log-likelihood in $\text{poly}(n,d,1/\epsilon)$ time. Both approaches are stochastic and rely on the work of \cite{LogconcaveSampling-07} to sample from $\varphi_n$ over the convex body $C_n$. \newb{Regarding the computational efficiency of the latter approach,  \cite{Hit-and-Run-Implementation-Lovasz-2012} noted that they "could not experiment with other convex bodies than cubes, because the oracle describing the convex bodies took too long to run". 
Since neither \cite{axelrod2018} nor \cite{diakonikolas2018} provide an implementation of their novel approaches, a fair and competitive evaluation has to be left for future work.}
}


%
\begin{figure}
\centering 
\subfloat[][$n$-gon with $n=25$]{\includegraphics[width=0.31\textwidth]{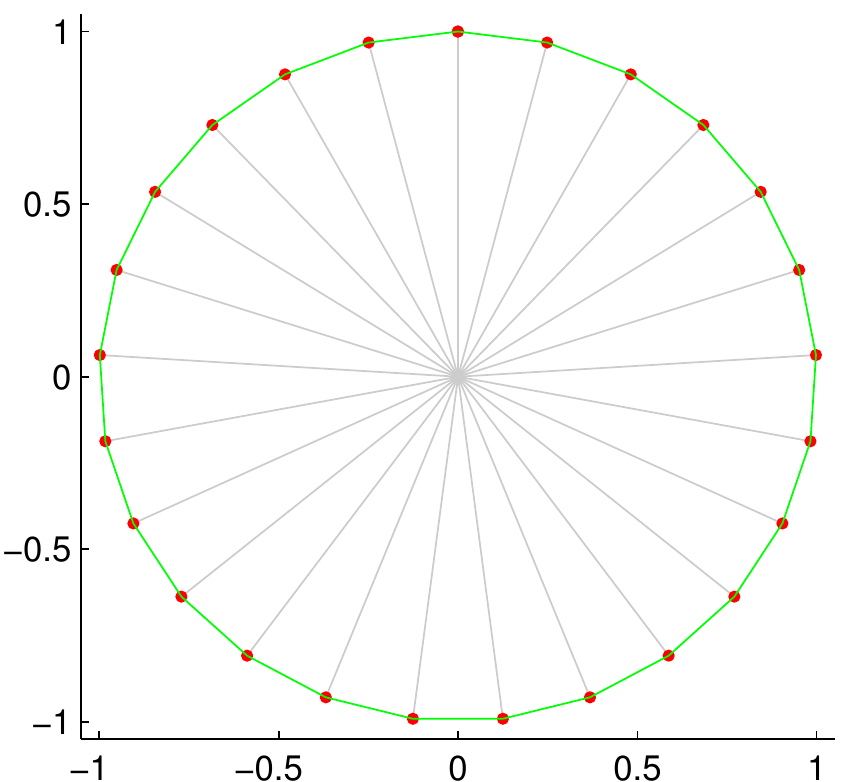}}
\hfill
\subfloat[][\centering \footnotesize{\cite{cule2010}, \par $l(\theta) = -29.4109$, $N_{n,d}=23$}]{\includegraphics[width=0.31\textwidth]{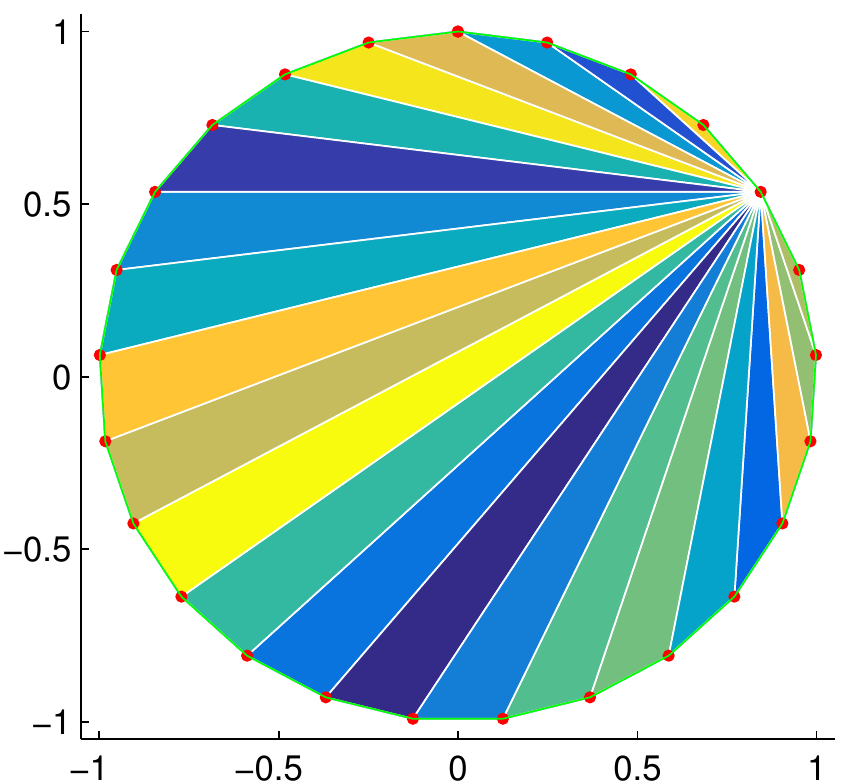}}
\hfill
\subfloat[][\centering \footnotesize{Our estimate, \par $l(\theta) = -29.4109$, $N_{n,d}=1$}]{\includegraphics[width=0.31\textwidth]{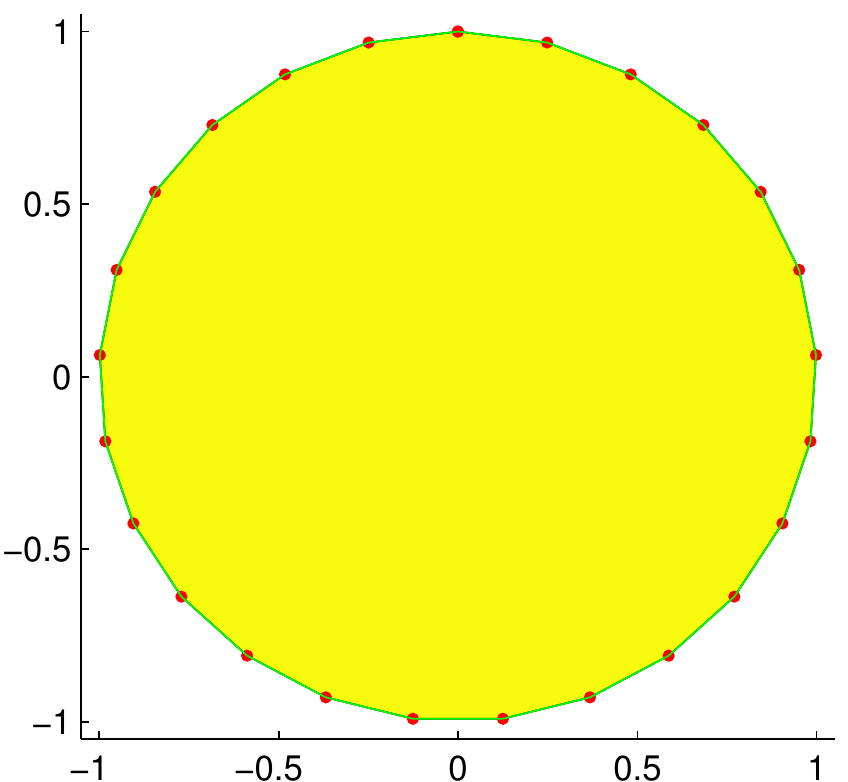}}
\caption{Example due to D.~Schuhmacher from the discussion in the appendix of \cite{cule2010} regarding computational efficiency: (a) n=25 data points $\mc{X}_{n}$ form a $n$-gon. Due to the symmetry of $\mc{X}_{n}$, the density estimate $\hat{f}_n$ is the uniform density (not explicitly shown here; both the approach \cite{cule2010} and our approach return this estimate, as the equal log-likelihood values $l(\theta)$ \eqref{eq:total-log-likelihood} demonstrate). It is clear that this uniform density can be represented by \textit{a single} hyperplane, $N_{n,d}=1$, 
that our approach correctly finds (panel (c)). In contrast, the approach of Cule et al.~(b) relies on a \textit{triangulation} of $C_n$, which leads to a more involved density parameter estimation problem: $N_{n,d}=23$ affine function parameters. This gap of complexity increases considerably with larger numbers $n$ of data points and dimension $d$ of the data space.
}
\label{fig:n-gon}
\end{figure}

\subsection{Contribution, Organization}
This preceding discussion motivated us to address the two shortcomings (i), (ii) raised above as follows. 
\begin{enumerate}[(1)]
\item We consider the representation \eqref{eq:def-Phi} of
$\hat\vphi_{n}$ and adopt a \textit{smooth} approximation of the non-smooth $\max$-operation. While the resulting modification of \eqref{eq:def-J} no longer is convex, numerical methods can be applied that are orders of magnitude more efficient than the subgradient based iterative schemes of \cite{cule2010}. Furthermore, we exploit the fact that the smoothness of the approximation can be controlled by a single parameter $\gamma$: While we utilize strong smoothing to obtain an initial parameter vector, the subsequent optimization is carried out with minimal smoothing.

\item
Rather than optimizing \textit{all} parameters of \eqref{eq:vphi-max}, we apply a threshold criterion in order to drop `inactive' hyperplanes, since the optimal estimate $\hat\vphi_{n}$ can be expected to be defined by a small subset of them, as discussed above. This measure speeds up the computations too without essentially compromising the accuracy of the resulting density estimator $\hat f_{n}$. Moreover, unlike the approach of \cite{cule2010}, we do not restrict polyhedral subsets $C_{n,i}$ to simplices. Figure \ref{fig:n-gon} shows a somewhat extreme academical example in order to  illustrate these points.
\end{enumerate}
Due to the non-convexity of our objective function, we cannot guarantee that our approach determines the maximum-likelihood density estimate for \textit{every} problem instance, as does the approach of \cite{cule2010}. This was the case, however, in a comprehensive series of numerical experiments indicate, that we report below. In particular, log-concave density estimation for large sample sets and for higher dimensions becomes computationally tractable.

Our paper is organized as follows. We present our approach in Section \ref{sec:method} and discuss  details of the algorithm and its implementation. 
In Section \ref{sec:results}, we report extensive numerical results up to dimension $d=6$ using sample sizes in the range $n \in [10^{2},10^{5}]$. In the univariate case $d=1$, our method is on par with the active set approach of \cite{duembgen2007} regarding both runtime and accuracy. This method is not applicable to higher dimensions, however. In such cases, $d \in \{2,\dotsc,6\}$, our method is as accurate as the algorithm of \cite{cule2010} but orders of magnitude more efficient. For example, for $d=2$ and $n=10.000$ samples, the algorithm of Cule et al.~takes $4.6$ hours whereas our algorithm terminates after $0.5$ seconds. For $d=6$ and $n=1.000$ samples, the algorithm of Cule et al.~takes about $10$ hours, whereas our algorithm terminates after $5$ minutes.

An implementation of our approach is publicly available as software R package \texttt{fmlogcondens} \citep{fmlogcondens} on CRAN.

\section{Approach}
\label{sec:method}

We define in Section \ref{sec:Objective-Function} the objective function as smooth approximation of the negative log-likelihood. The subsequent sections discuss how the parameter values of the log-concave density estimate are determined by numerical optimization. The overall structure of the approach is summarized as Algorithm \ref{alg-1} on page \pageref{alg-1}.  

\subsection{Objective Function}
\label{sec:Objective-Function}
We rewrite the
negative log-likelihood functional \eqref{eq:def-J} in the form
\begin{subequations}\label{eq:def-L-theta}
\begin{align}
L(\theta) &:= \frac{1}{n} \sum_{i=1}^{n} \vphi_{n}(x_{i}) + \int_{C_{n}} \exp\big(-\vphi_{n}(x)\big)\dd{x}, \\
\theta &:= \big\{(a_{1},b_{1}),\dotsc,(a_{N_{n,d}},b_{N_{n,d}})\big\},
\end{align}
\end{subequations}
where $\vphi_{n}$ and all $\vphi_{i,n}$ have the form \eqref{eq:vphi-affine} and \eqref{eq:vphi-max}, respectively, and $\theta$ collects all parameters that determine $\vphi_{n}$. We define the log-concave density estimate \eqref{eq:def-hat-f} in terms of the function 
\begin{equation}
\hat\vphi_{n} = \vphi_{n}|_{\theta=\hat\theta} \colon \quad
\hat\theta\;\text{locally minimizes}\; L(\theta).
\end{equation}
Our next step is to smoothly approximate the representation \eqref{eq:vphi-max} of $\vphi_{n}$. Using the convex log-exponential function
\begin{equation}
\logexp \colon \R^{d} \to \R,\qquad
x \mapsto \logexp(x) := \log\Big(\sum_{i=1}^{d} e^{x_{i}}\Big)
\end{equation}
we introduce a \textit{smoothing parameter} $\gamma>0$ and define the rescaled smooth convex function
\begin{equation}
\label{eq:logexp}
\logexp_{\gamma} \colon \R^{d} \to \R,\qquad
x \mapsto \logexp_{\gamma}(x) := \gamma \logexp\Big(\frac{x}{\gamma}\Big)
= \gamma \log\bigg(\sum_{i=1}^{d} \exp\Big(\frac{x_{i}}{\gamma}\Big)\bigg),
\end{equation}
that \textit{uniformly} approximates the non-smooth $\max$-operation \cite[Example~1.30]{rockafellar2009} in the following sense:
\begin{equation}\label{eq:logexp-bounds}
\logexp_{\gamma}(x) - \gamma \log d 
\leq \max_{i=1,\dotsc,d}\{x_{1},\dotsc,x_{d}\}
\leq \logexp_{\gamma}(x),\qquad 
\forall x \in \R^{d}.
\end{equation}
Utilizing this function, we define in view of \eqref{eq:vphi-max} the smooth approximation
\begin{equation}
\label{eq:vphi-smooth-approx}
\vphi_{n,\gamma}(x) := \begin{cases}
\logexp_{\gamma}\big(\vphi_{1,n}(x),\dotsc,\vphi_{N_{n,d},n}(x)\big), & x \in C_{n}, \\
\infty, & x \not\in C_{n},
\end{cases}
\end{equation}
and in turn the smooth approximation of the objective function \eqref{eq:def-L-theta}
\begin{equation}\label{eq:def-L-theta-gamma}
L_{\gamma}(\theta) := \frac{1}{n} \sum_{i=1}^{n} \vphi_{n,\gamma}(x_{i}) + \int_{C_{n}} \exp\big(-\vphi_{n,\gamma}(x)\big)\dd{x}.
\end{equation}
We point out that by virtue of \eqref{eq:logexp-bounds}, we have
\begin{equation}
\forall x \in C_{n},\qquad
0 \leq \vphi_{n,\gamma}(x)-\vphi_{n}(x) \leq \gamma \log d \;\to 0\quad\text{for}\quad \gamma \to 0
\end{equation}
and consequently, by continuity, 
\begin{equation}
L_{\gamma}(\theta) \to L(\theta)\quad\text{for}\quad \gamma \to 0.
\end{equation}
\subsection{Numerical Optimization}\label{sec:optimization}
We apply an established, memory-efficient quasi-Newton method known as L-BFGS in the literature \citep{nocedal2006}, to compute a sequence 
\begin{equation}
\big(\theta^{(k)}\big)_{k \geq 1}
\end{equation}
of parameter values that converges to a local minimum $\hat\theta$ of the objective function \eqref{eq:def-L-theta-gamma}. A key aspect of this iterative procedure is to maintain at each step $k$ an approximation $H^{(k)}$ of the inverse Hessian $\big(\nabla^{2} L_{\gamma}(\theta^{(k)})\big)^{-1}$ 
of the objective function \eqref{eq:def-L-theta-gamma}, in terms of a few gradients $\nabla L_{\gamma}(\theta^{(k')})$ evaluated and stored at preceding iterative steps $k' < k$. This avoids to handle directly the Hessian of size $(\dim \theta)^{2} = \big((d+1) N_{n,d}\big)^{2}$ and hence enables to cope with much larger problem sizes. 

The basic update steps with search direction $p^{(k)}$ and step size $\lambda_{k}$ read
\begin{equation}\label{eq:lambda-update}
\theta^{(k+1)} = \theta^{(k)} + \lambda_{k} p^{(k)},\qquad
p^{(k)} = -H^{(k)} \nabla L_{\gamma}(\theta^{(k)}).
\end{equation}
The stepsize $\lambda_{k}$ is determined by backtracking line search. \new{More specifically, we select the largest $\lambda_k$ in the set $\{1, p, p^2, \ldots \}, p=0.1,$ such that the condition
\begin{equation}
L_{\gamma}(\theta^{(k)} + \lambda_k p^{(k)}) -  L_{\gamma}(\theta^{(k)}) \leq \sigma \lambda_k (p^{(k)})^T \nabla L_{\gamma}(\theta^{(k)}) 
\end{equation}
holds. We chose $\sigma = 10^{-2}$, meaning we accept a decrease in $L_{\gamma}$ by 1\% of the prediction based on the linear extrapolation.
}

\new{Now, instead of computing $H^{k}$ anew in every iteration, it is merely updated to account for the curvature measured in the most recent step. Given a new iterate $\theta^{(k+1)}$, the update for $H^{(k)}$ in the BFGS approach is \citep[Chap.~6.1]{nocedal2006}
\begin{equation}\label{eq:BFGS-update}
H^{(k+1)} = (V^{(k)})^T H^{(k)} V^{(k)} + \rho^{(k)}s^{(k)} (s^{(k)})^T,
\end{equation}
where 
\begin{equation*}
\rho^{(k)} = \frac{1}{(y^{(k)})^T s^{(k)}}, \qquad V^{(k)} = I - \rho^{(k)} y^{(k)
} (s^{(k)})^T, 
\end{equation*}
and 
\begin{equation*}
s^{(k)} = \theta^{(k+1)} - \theta^{(k)}, \qquad
y^{(k)} = \nabla L_{\gamma}(\theta^{(k+1)}) - \nabla L_{\gamma}(\theta^{(k)}).
\end{equation*}
This update has the property, that if $H^{(k)}$ is positive definite and the \textit{curvature condition} 
\begin{equation}\label{eq:curvature-condition}
(y^{(k)})^T s^{(k)} > 0
\end{equation}
is fulfilled, then $H^{(k+1)}$ is also positive definite, which in turn guarantees that the step $p^{(k+1)}$ \eqref{eq:lambda-update} is a descent direction.}

\new{While \eqref{eq:curvature-condition} automatically holds in the convex case, this property has to be enforced explicitly for non-convex objective functions.}
\cite{Non-Convex-BFGS-2001}, therefore,  proposed the following modification of $y^{(k)}$:
\begin{equation}
\tilde{y}^{(k)} = y^{(k)} + t_k s^{(k)}, \qquad t_k = \| \nabla L_{\gamma}(\theta^{(k)}) \| + \max\left\{ -\frac{(y^{(k)})^T s^{(k)}}{\| s^{(k)} \|^2}, 0 \right\},
\end{equation}
which fulfills \eqref{eq:curvature-condition} since $(\tilde{y}^{(k)})^T s^{(k)} \geq \| \nabla L_{\gamma}(\theta^{(k)}) \| \| s^{(k)} \|^2 > 0$. Thus using $\tilde{y}^{(k)}$ in \eqref{eq:BFGS-update} guarantees the positive definiteness of $H^{(k+1)}$. \new{See \citet[Thm.~5.1]{Non-Convex-BFGS-2001} for a proof of convergence.}

\new{Storing $H^{(k)}$ in memory quickly becomes prohibitive with growing $\dim (\theta)$. This is addressed by \textit{limited-memory} BFGS (L-BFGS)  by only storing the $m$ most recent vectors $(y^{(k)}, s^{(k)})$, representing $H^{(k)}$ implicitly. At every iteration $p^{(k)}$ \eqref{eq:lambda-update} is directly calculated by recursively applying formula \eqref{eq:BFGS-update}, see~\cite[Ch.~7.2]{nocedal2006}. As a result, this approximation of $H^{(k)}$ only requires the curvature information of the last $m$ steps. We set $m=40$ to obtain a reasonably accurate approximation.}

\subsection{Numerical Integration} The numerical optimization steps of Section \ref{sec:optimization} require the accurate integration of smooth functions over $C_{n}$, due to \eqref{eq:def-L-theta-gamma}.

We examined various numerical integration schemes: Sparse grid approaches \citep{Sparse-Grids-2004-Bungertz} utilize truncated tensor products of one-dimensional quadrature rules and scale well with the dimension $d$. But they are not practical for integrating over non-quadrilateral domains \citep[Sec.~5.3]{Sparse-Grids-2004-Bungertz}, an observation confirmed by our experiments. Another family of approaches are Monte-Carlo methods based on random-walks \citep{Hit-And-Run-2006-Lovasz,Hit-And-Run-Rudolf-2012}, that specifically address the problem of integrating a log-concave density~$f$ over a convex body. Nevertheless, experimental results \citep{Hit-and-Run-Implementation-Lovasz-2012} raised doubts about their efficiency, and we did not further pursue this type of approach.

\begin{figure}
\centerline{\includegraphics[width=0.45\textwidth]{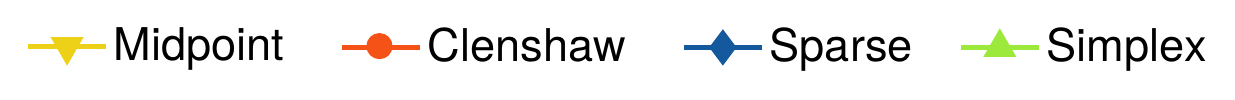}}
\centering 
\subfloat[][Clenshaw-Curtis Quadrature]{\includegraphics[width=0.22\textwidth]{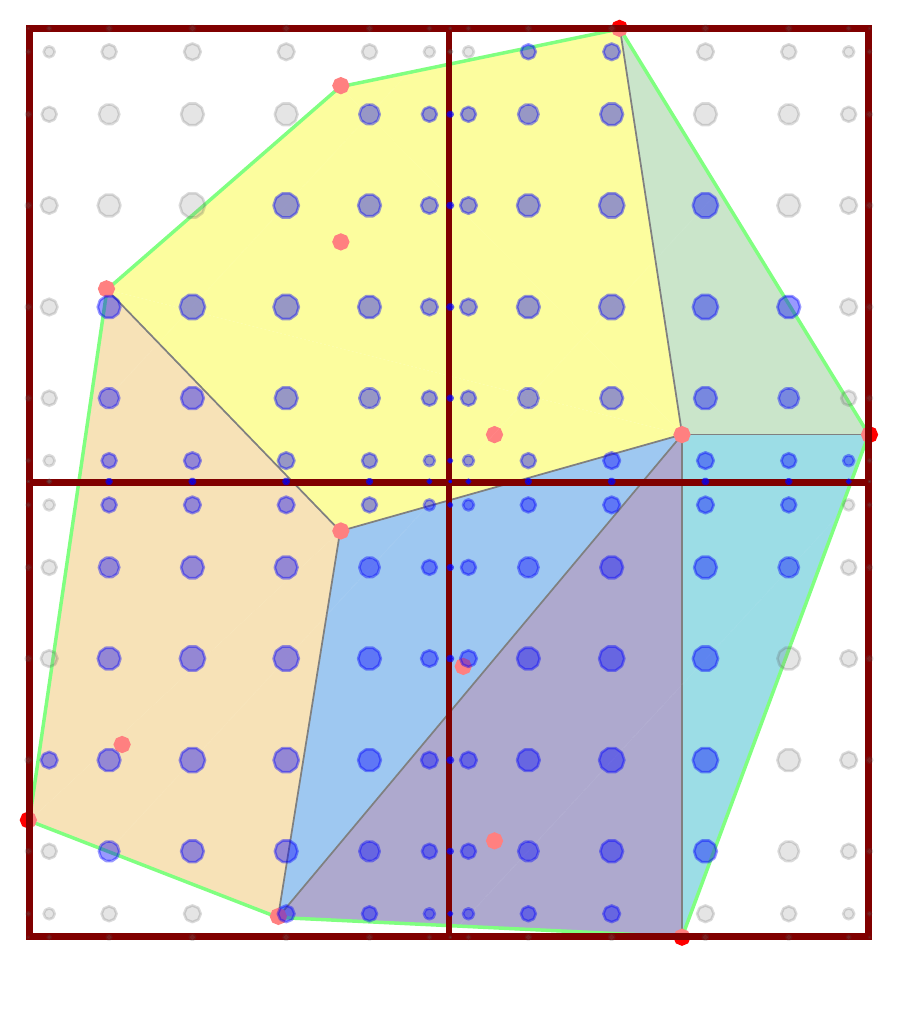}}
\hfill
\subfloat[][Quadrature rule for simplices]{\includegraphics[width=0.22\textwidth]{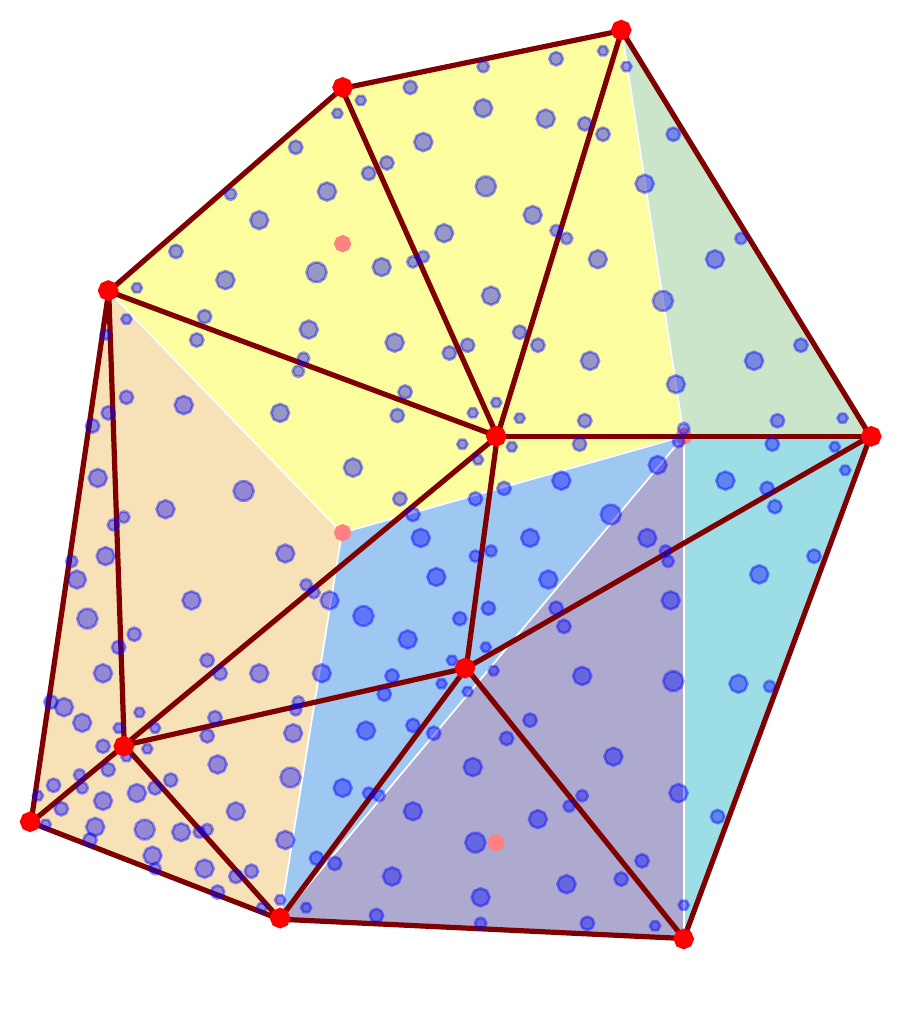}}
\hfill
\subfloat[][$d=2, n=1000$]{\includegraphics[width=0.24\textwidth]{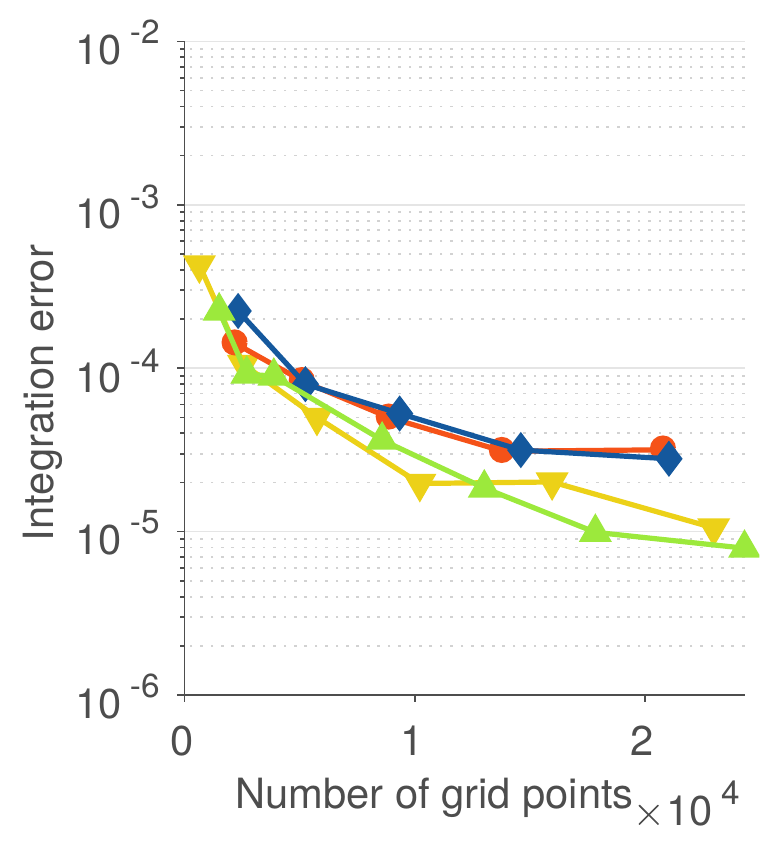}}
\hfill
\subfloat[][$d=4, n=1000$]{\includegraphics[width=0.24\textwidth]{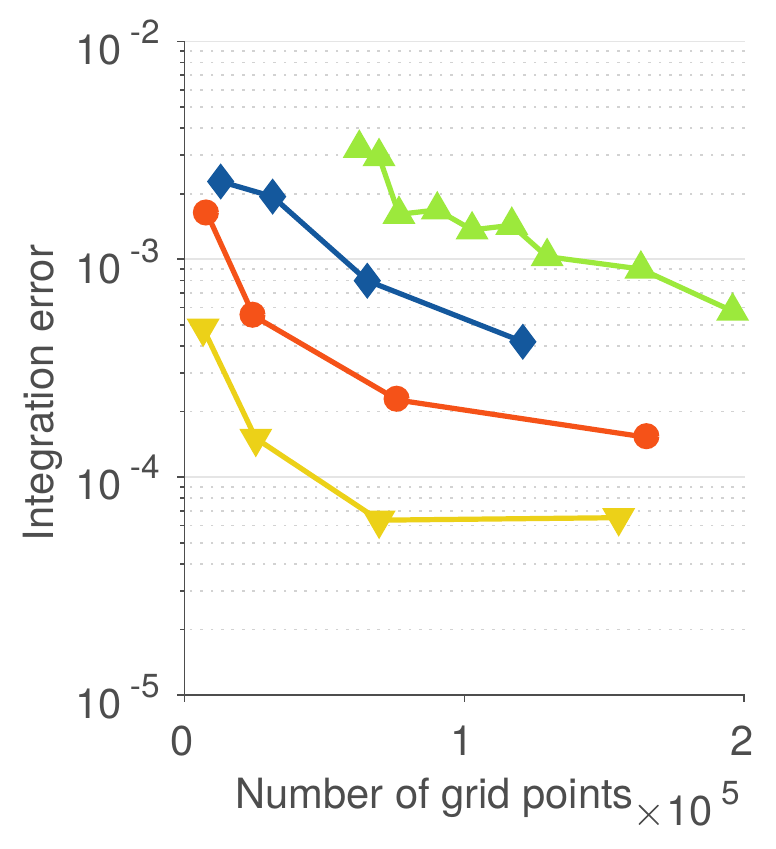}}
\hfill
\caption{(a-b) Two integration schemes for the  dataset from Figure~\ref{fig:tent-function} illustrate difficulties that arise for convex integration areas: Integration schemes (a) designed for cubic integration areas lose accuracy when truncated outside the convex integration domain (gray dots). Schemes (b) suited for simplical integration domains degrade when the simplices are not well aligned to the polyhedral subdivision of $C_n$ induced by $\vphi_n(x)$ \eqref{eq:vphi-affine}. Higher dimensions $d$ aggravate these effects. (c-d) Simple Riemann sums using the midpoint rule and uniform weights performed best in our experiments, as they do not assume any specific integration area. Simplex based schemes worked only well for small $n$ and $d$.}
\label{fig:integration}
\end{figure}

In addition, we examined various dense integrations schemes for the hypercube (simple Newton-Cotes schemes and Clenshaw-Curtis quadrature) as well as schemes tailored to simplical integration domains, e.g.~\cite{grundmann1978}. Again, regarding the quadrature rules designed for the hypercube, the need to truncate them outside convex subsets (illustrated in Figure~\ref{fig:integration} (a)) had a negative impact on integration accuracy. On the other hand, the simplex-based schemes only worked well if the randomly chosen simplical decomposition of $C_n$ for the integration resembled the decomposition \eqref{eq:Cn-decomposition} of $\hat{\vphi}_n$. This was only the case for small $d$ and $n$, however. Figure~\ref{fig:integration} (b) illustrates a typical case of misalignment.

Overall, simple Riemann sums with uniform integration weights performed best in our experiments, because the influence of the shape of the integration domain is minor (Figure~\ref{fig:integration} (c-d)). For future reference, we denote the integration grid by
\begin{equation}
\mc{Z}_m =\{z_1, \ldots, z_m\} \subset \R^d,
\end{equation}
and the uniform integration weights by $\Delta$. Accordingly, the numerical approximation of the integral of the objective \eqref{eq:def-L-theta-gamma} reads
\begin{equation}
\label{eq:numerical-integration}
\int_{C_{n}} \exp\big(-\vphi_{n,\gamma}(x)\big)\dd{x} \approx \Delta \sum_{i=1}^m \exp\big(-\vphi_{n,\gamma}(z_i)\big).
\end{equation}

\new{While naively evaluating \eqref{eq:numerical-integration} for the combination of all hyperplanes and grid points would quickly become intractable, we point out that the impact of each hyperplane $(a_i, b_i)$ is close to zero for most grid points. This fact combined with further plausible measures renders the integration task tractable even for larger dimensions. See \ref{sec:implementation} for details and discussion.}

\begin{figure}
\subfloat[Log-likelihood]{\includegraphics[width=0.31\textwidth]{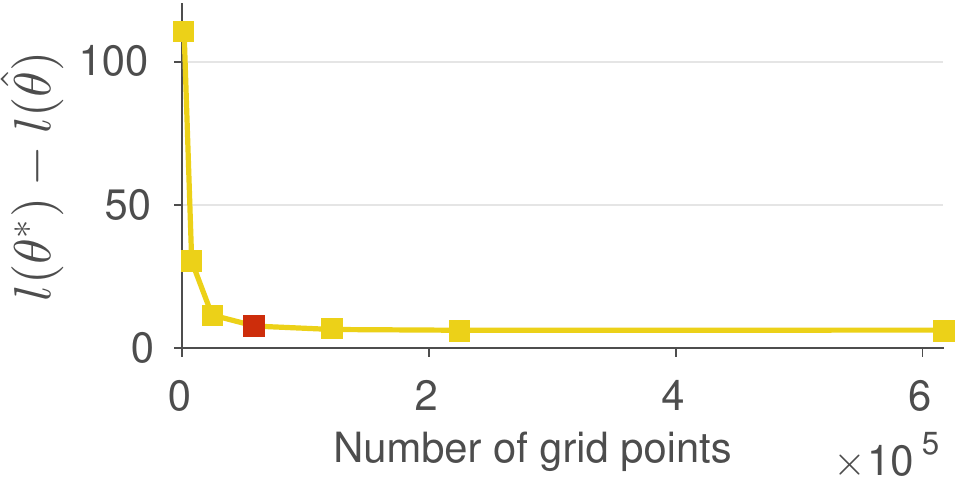}}
\hfill
\subfloat[Runtime]{\includegraphics[width=0.31\textwidth]{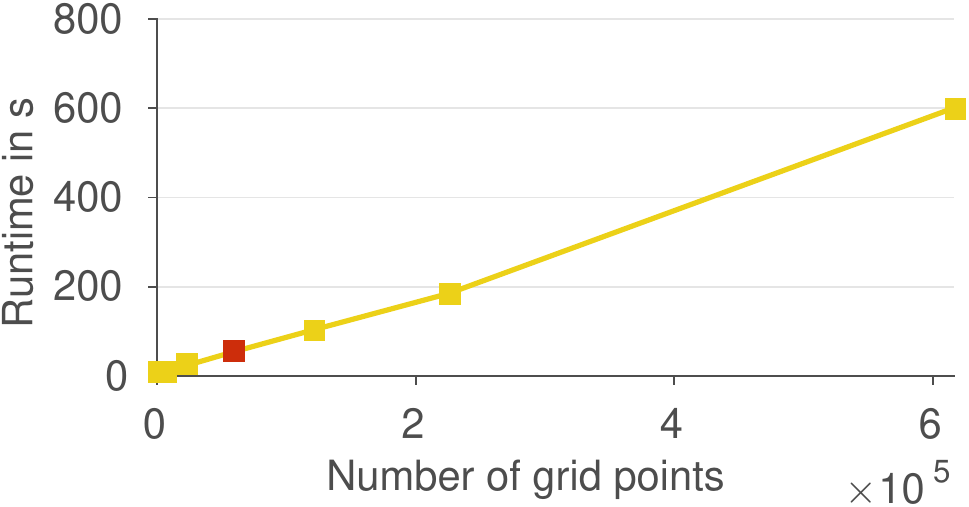}}
\hfill
\subfloat[Number of hyperplanes]{\includegraphics[width=0.31\textwidth]{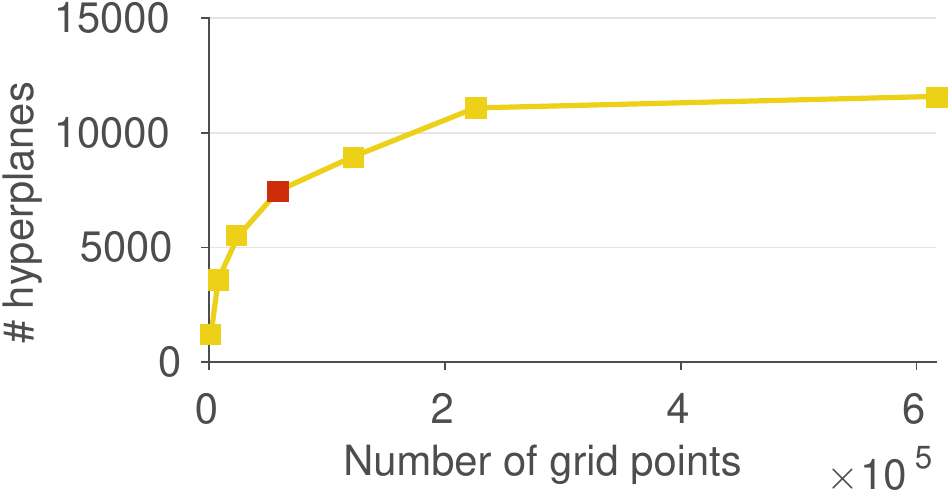}}
\caption{\newb{Effect of the grid density (number of grid points of $\mathcal{Z}_m$) versus (a) quality, (b) runtime and (c) complexity of the solution for a sample of $n=5000$ points in $\R^4$. As the log-likelihood converges rapidly along with the number of hyperplanes, the runtime increases linearly with the number of grid points. The red marked square defines the density used in our implementation, a trade-off between accuracy and runtime.}}
\label{fig:grid-density}
\end{figure}

\new{Regarding the \textit{density} of the integration grid, we traded off accuracy against computational complexity. Figure \ref{fig:grid-density} (a) illustrates, for a sample of 5000 points in $\R^4$, how the accuracy improves with increasing grid density and finally converges to a solution close to the optimum. As expected, the runtime grows linearly with the number of grid points (Figure \ref{fig:grid-density} (b)). In general we found the ratio of grid points to number of hyperplanes (Figure \ref{fig:grid-density} (c)) to be a good performance indicator. Keeping this ratio above 3 yielded good results, and we set a minimal number of grid points based on the expected number of hyperplanes for each dimension $d$ (except for 6-D, where we chose a lower ratio for performance reasons). The red square indicates our choice for 4-D.}

\subsection{Initialization}
\label{sec:initialization}
Initialization of $\theta$ plays a crucial role due to the non-convexity of $L_{\gamma}(\theta)$. We examined two different approaches: The \textit{first approach} is based on a kernel density estimate $f_{\text{kernel}}(x)$ as in \cite{cule2010}, using a multivariate normal kernel with a diagonal bandwidth matrix $M$ with entries
\begin{equation*}
M_{jj} = \sigma_j n^{-1/(d+4)}, \qquad j = 1, \ldots d,
\end{equation*}
where $\sigma_j$ is the standard deviation of $\mc{X}_n$ in dimension $j$. Setting $y_i = \log f_{\text{kernel}}(x_i)$ for $i=1, \ldots, n$, we compute a simplical decomposition of $C_{n}$ induced by the upper convex hull of $(X,y)$, using the popular quickhull algorithm \citep{barber1996quickhull}. The simplical decomposition combined with $y$ then yields an initial set of hyperplane parameters $\theta^{(0)}$, one for each simplex $C_{n,i}$.
%

As for the \textit{second approach}, we randomly initialize a small number of hyperplanes and optimize $L_{\gamma}(\theta)$ with $\gamma = 1$. The rational behind this is that since $\gamma$ governs the degree of non-convexity and smoothness of $L_{\gamma}(\theta)$, 
its optimization is less involved than for smaller $\gamma$. Having found the optimal log-concave density for $\gamma=1$, we evaluate $y_i$ for all $x_i$ and proceed as described above (first approach) to obtain~$\theta^{(0)}$. \new{Regarding the specific choice for $\gamma$, experiments showed that initializations with $\gamma = 1$ yielded superior results compared to other initial values of $\gamma$, thus offering the ``best'' trade-off between smoothness of the objective and initial accuracy of the max approximation.}

Except for small datasets, in general the second initialization performs better. In practice, we calculate both and select the one with smaller $L_{\gamma}(\theta^{(0)})$.

\subsection{Pruning Inactive Hyperplanes} 
\label{sec:pruning}
Both initializations produce a very large set of hyperplanes based on a simplical decomposition of $C_n$, with one hyperplane per simplex. During the optimization, hyperplanes may become inactive. Inactivity of some hyperplane $(a_j,b_j)$ in the light of \eqref{eq:vphi-affine}  means that there exists no $x \in C_n$ for which $\hat{\vphi}_n(x) = \la a_j, x \ra + b_j$. In terms of our smooth approximation $\varphi_{n,\gamma}(x)$ \eqref{eq:vphi-smooth-approx}, every hyperplane contributes due to $\exp(x) > 0, \;\forall x \in \R$, albeit the contribution may be very close to $0$. We therefore resort to the following definition of inactivity using our integration grid:
\begin{equation}
\sum_{i=1}^m \frac{\exp(\gamma^{-1}(\la a_j, z_i \ra + b_j))}{\sum_{k=1}^{N_{n,d}} \exp(\gamma^{-1}(\la a_k, z_i \ra + b_k))} \leq \vartheta.
\label{eq:inactivity-constraint}
\end{equation}
After each update of $\theta^{(k)}$ we remove all hyperplanes that satisfy \eqref{eq:inactivity-constraint} with $\vartheta = 10^{-3}$, which corresponds to a total contribution of less than $10^{-3}$ grid points. We chose this criterion after observing that hyperplanes usually remained inactive once they lost support on the integration grid, due to their very small  contribution to the objective function gradient.

\begin{figure}
\centering 
\subfloat[][$\theta^{(0)}, N_{n,d} = 99$]{\includegraphics[width=0.27\textwidth]{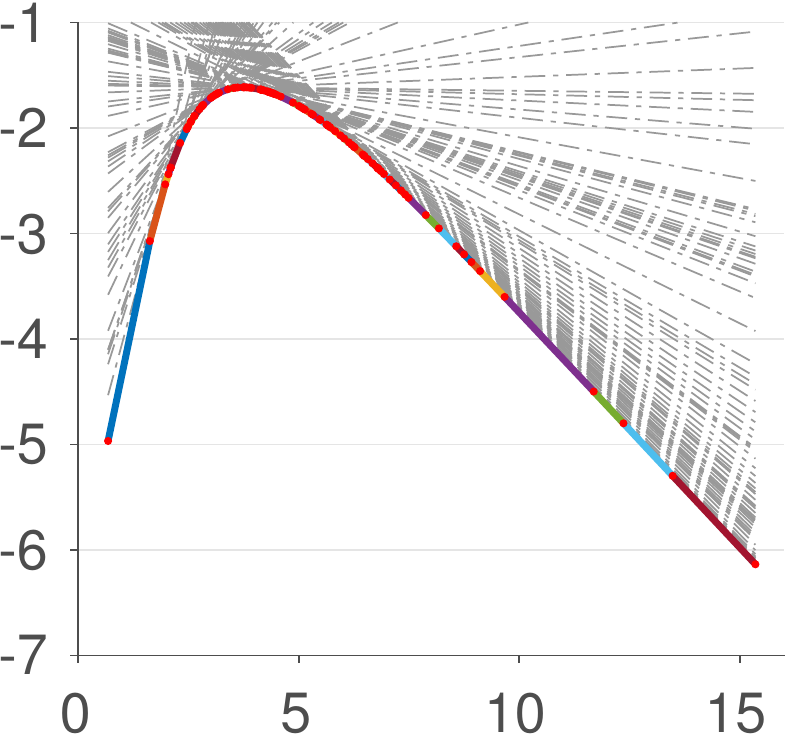}}
\hspace{0.5cm}
\subfloat[][$\theta^{(5)}, N_{n,d} = 45$]{\includegraphics[width=0.27\textwidth]{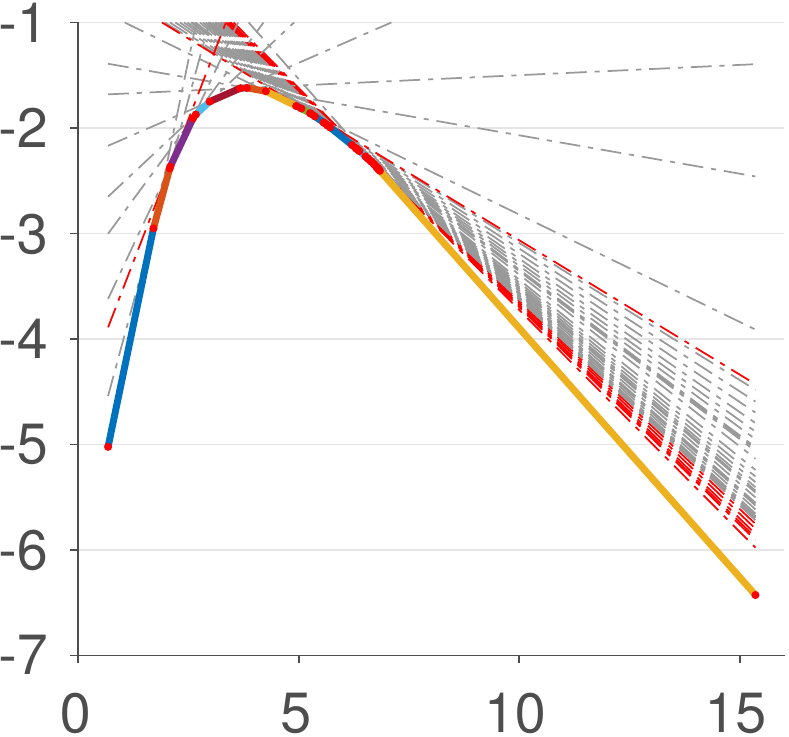}}
\hspace{0.5cm}
\subfloat[][$\theta^{(9)}, N_{n,d} = 26$]{\includegraphics[width=0.27\textwidth]{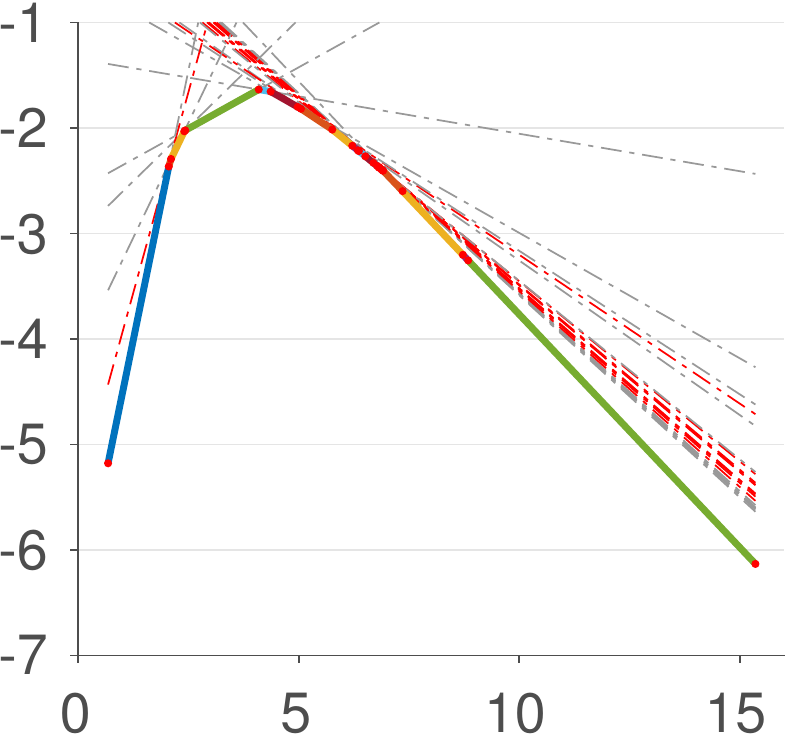}}

\subfloat[][\centering Final solution \par $\theta^{(\text{final})}, N_{n,d} = 5$]{\includegraphics[width=0.27\textwidth]{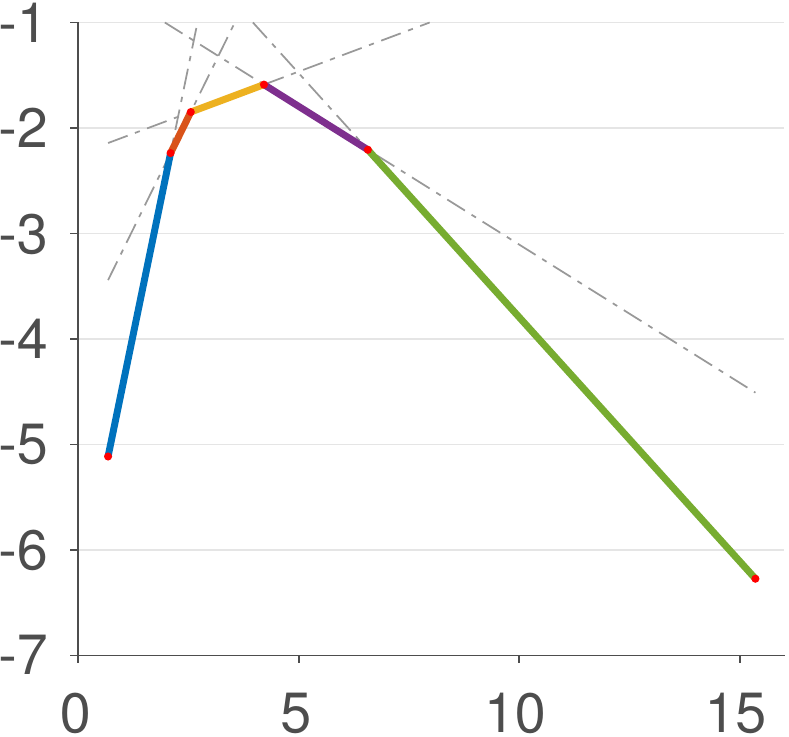}}
\hspace{0.5cm}
\subfloat[][\centering Optimal solution, $N_{n,d}=5$, \par \cite{duembgen2007}]{\includegraphics[width=0.27\textwidth]{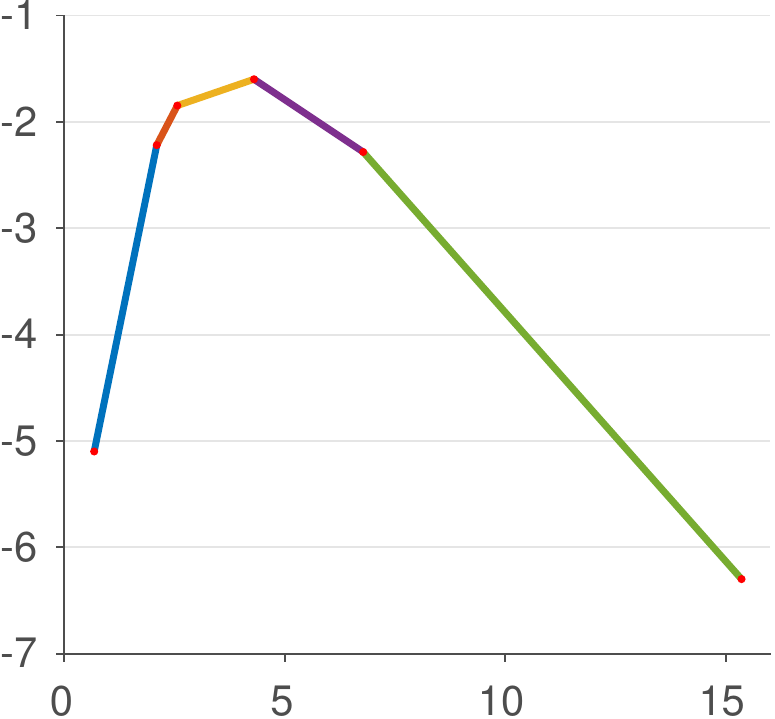}}
\hspace{0.5cm}
\subfloat[][\centering Optimal solution, $N_{n,d}=20$, \par \cite{cule2010}]{\includegraphics[width=0.27\textwidth]{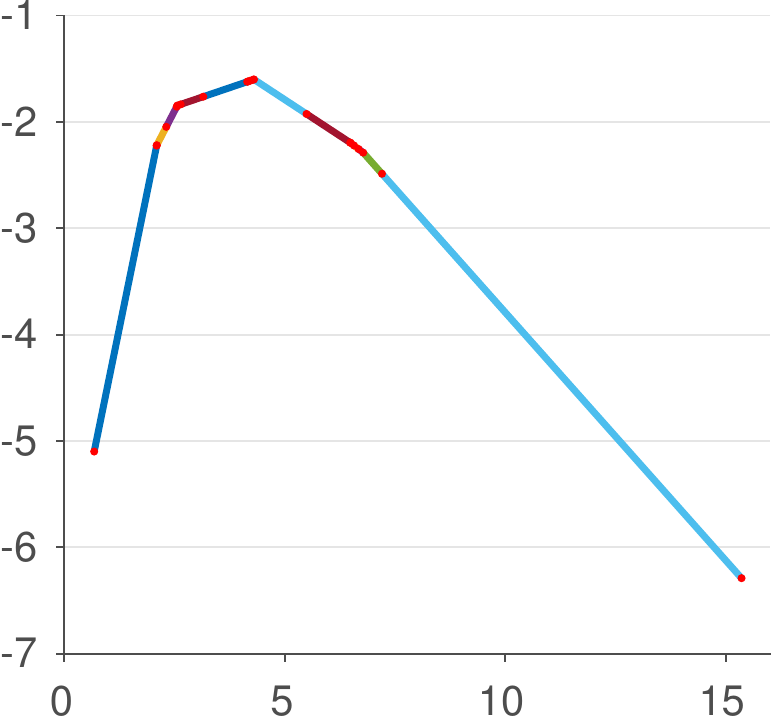}}
\caption{(a)-(d) Several steps of the optimization process for a sample of 100 points in $\R$ drawn from a gamma distribution $\text{Gamma}(\alpha=5, \beta=1)$. Hyperplanes to be dropped in the next step are drawn in red. (a) The initial density is represented by 99 hyperplanes, one for each simplex of the simplical decomposition of $C_n$ and based on the smooth max-approximation $\varphi_{n,\gamma=1}(x)$ (see Section \ref{sec:termination}). Plots (b) and (c) show the transition towards the optimal final shape composed of $N_{n,d}=5$ hyperplanes (d). For comparison the non-sparse solution of \cite{cule2010} (f) and in (e) the optimal solution (only available in 1-D) with a \textit{minimal} representation by \cite{duembgen2007} which is identical to the solution returned by our approach.}
\label{fig:pruning}
\end{figure}

Figure~\ref{fig:pruning} visualizes several intermediate steps during an optimization process for $d=1$, together with the shrinking set of hyperplanes. Plots (d)-(f) show the effectiveness of this scheme, since we arrive at the same parametrization as the approach of \cite{duembgen2007}, which -- provided $d=1$ -- finds the \textit{minimal} representation for $\hat{\vphi}_n(x)$ in $\R$.

\subsection{Termination}
\label{sec:termination}
To determine convergence at each iteration, we check if the following inequalities are satisfied:
\begin{subequations}
\label{eq:termination-constraints}
\begin{align}
| 1 - \Delta \sum_{i=1}^{m} \exp\big(-\hat{\varphi}_{n,\gamma}(z_i) \big) | &\leq \epsilon, \\
| L_{\gamma}(\theta^{(k+1)}) - L_{\gamma}(\theta^{(k)}) | & \leq \delta,
\end{align}
\end{subequations}
where we use $\epsilon = 10^{-3}$ and $\delta = 10^{-7}$. The first criterion asserts that the current density estimate $\hat f_{n} = \exp(-\hat\vphi_{n,\gamma})$ satisfies $\int \hat f_{n} \dd{x} \geq 1-\veps$. Then second condition detects when the decrease of the objective function becomes negligible. We denote the final parameter vector by $\theta^{(\text{final})}$.

\subsection{Exact Normalization}
\label{sec:normalization}
As a final step, after convergence of the optimization algorithm, we normalize the estimated density using \textit{exact} integration and the non-smoothed representation \eqref{eq:vphi-max} of $\hat \vphi_{n}$, which may be seen as setting $\gamma=0$ in \eqref{eq:vphi-smooth-approx}.
Setting $y_i = \hat{\varphi}_{n,\gamma}(x_i)$ for all $x_i \in \mc{X}_n$, we again use \texttt{qhull} \citep{barber1996quickhull} to obtain a triangulation of $C_n$ and calculate a hyperplane $\hat{\vphi}_{i,n}$ for every simplex $C_{n,i}$. We then split the integral over $C_n$ into separate integrals over simplices $C_{n,i}$ and denote the result by $\lambda$:
\begin{equation}
\label{eq:analytical-integral}
\int_{C_n} \exp(-\hat{\vphi}_n(x)) \dd{x} = \sum_{i=1}^{N_{n,d}} \int_{C_{n,i}} \exp( - \hat{\vphi}_{i,n}(x)) \dd{x} := \lambda
\end{equation}
We make use of Lemma \ref{lemma:integral-cule} to evaluate \eqref{eq:analytical-integral} exactly. 

The value of $\lambda$ is close to $1$ but \textit{not equal} to $1$. 
We therefore add the same offset parameter $\delta$ to every hyperplane $\hat{\vphi}_{i,n}$, to obtain
\begin{equation}
\label{eq:modified-hyperplane}
\tilde{\vphi}_{i,n}(x) := \hat{\vphi}_{i,n}(x) + \delta = \la x, a_i \ra + b_i + \delta, \qquad i = 1, \ldots, N_{n,d}.
\end{equation}
Inserting $\tilde{\vphi}_{i,n}$ into \eqref{eq:exact-integral} shows that the integral for the modified hyperplanes \eqref{eq:modified-hyperplane} changes to $\exp(-\delta) \lambda$. Therefore, after setting $\delta = \log(\lambda)$, the final density estimate is $\hat f_{n} = \exp(-\tilde\vphi_{n})$ with $ \tilde\vphi_{n}|_{C_{n,i}} = \tilde{\vphi}_{i,n}$ given by \eqref{eq:modified-hyperplane}. We denote the corresponding \textit{dense} parameter vector by 
\begin{equation}
\label{eq:theta-hat}
\hat{\theta}.
\end{equation}

The normalization process eliminates the sparsity of the parametrization used for optimization, since it relies on a simplical decomposition of $C_n$, as does the approach of \cite{cule2010}. Nevertheless, the results of our approach are shown using the sparse parameterization $\theta^{(\text{final})}$, which is the essential characteristic of our approach causing the significant speed ups reported in Section~\ref{sec:results}, whereas the reported log-likelihood values are for $\hat{\theta}$.

\begin{algorithm}
\SetAlgoLined
\caption{Fast Log-Concave Density Estimation
\label{alg-1}}
\KwIn{$X$, parameters: $\gamma = 10^{-3}, \vartheta=10^{-3}, \epsilon = 10^{-3}, \delta=10^{-7}$}
\KwOut{Log-concave density estimate $\hat{f}_n$ parametrized by ${\theta}$ \eqref{eq:def-L-theta}.}
Find initial $\theta^{(0)}$ (Section \ref{sec:initialization})\;
\For{$k=1,2,\ldots$}{
Delete inactive hyperplanes from $\theta^{(k)}$ based on criterion \eqref{eq:inactivity-constraint}\; 
Compute the gradient $\nabla L_{\gamma}(\theta^{(k)})$ of the objective \eqref{eq:def-L-theta-gamma} using numerical integration\;
Find descent direction $p^{(k)}$ from the previous $m$ gradients vectors and step size $\lambda_k$ \eqref{eq:lambda-update} and update $\theta^{(k+1)}$\; 
\If {the termination criterion \eqref{eq:termination-constraints} holds,}{
	Denote final parameter vector by $\theta^{(\text{final})}$\;
	Quit for-loop\;
}
}
Switch from $\hat{\varphi}_{n,\gamma}$ to $\hat{\vphi}_{n}$ and perform exact normalization: \newb{$\theta^{(\text{final})} \rightarrow \hat{\theta}$} (Section \ref{sec:normalization})\;
\newb{\Return{$\hat{\theta}$ \eqref{eq:theta-hat}}}
\end{algorithm}

\section{Experiments}
\label{sec:results}
This section provides empirical evidence that our approach can find sparse solutions that are very close to the optimum and determined computationally using substantially less CPU runtime. Section \ref{sec:Qualitative-Evaluation} contrasts qualitative properties of our approach with the state of the art, using experiments in dimensions $d \in \{1,2\}$ along with illustrations. Quantitative results up to dimension $d=6$ are reported in Section \ref{sec:empirical-evaluation}. Finally, we extend our approach to the estimation of mixtures of log-concave densities in Section \ref{sec:Mixtures-Log-Concave}. 

\new{For all our experiments we used $\gamma = 10^{-3}$ for determining the accuracy of the smooth approximation $\varphi_{n,\gamma}$ \eqref{eq:vphi-smooth-approx} to the max function. This choice is a compromise between less accurate approximations (that is larger gammas) and more accurate but numerically unstable ones.} \newb{As noted in the previous section, qualitative results show $\theta^{(\text{final})}$, i.e. the sparse parametrization found during the optimization (Section \ref{sec:optimization} - \ref{sec:termination}), whereas quantitative results are reported in terms of $\hat{\theta}$, the \textit{dense} parametrization obtained by performing the final analytical normalization (Section \ref{sec:normalization}), and the non-smoothed representation \eqref{eq:vphi-max} of $\vphi_n$. The quality of our solutions is measured in terms of
\begin{equation}
\label{eq:total-log-likelihood}
l(\hat{\theta}) = \sum_{i=1}^n -\vphi_n(x_i),
\end{equation}
the total log-likelihood.}

\subsection{Qualitative Evaluation}\label{sec:Qualitative-Evaluation}
\begin{figure}
\centering 
\subfloat[][\centering \footnotesize{D\"{u}mbgen et al.~(\SI{0.23}{s}), \par $l(\hat{\theta}) = -683.33$, $N_{n,d}=9$}]
{\includegraphics[width=0.29\textwidth]{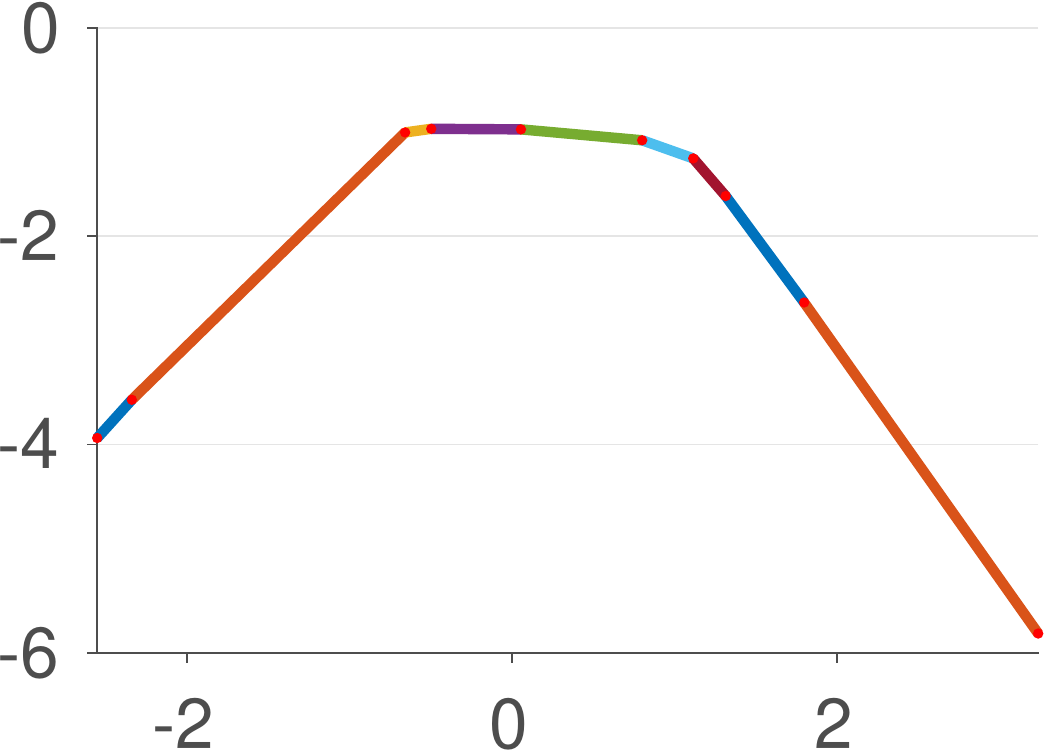}}
\hspace{0.5cm}
\subfloat[][\centering \footnotesize{Cule et al.~(\SI{3.26}{s}), \par $l(\hat{\theta}) = -683.34$, $N_{n,d}=45$}]
{\includegraphics[width=0.29\textwidth]{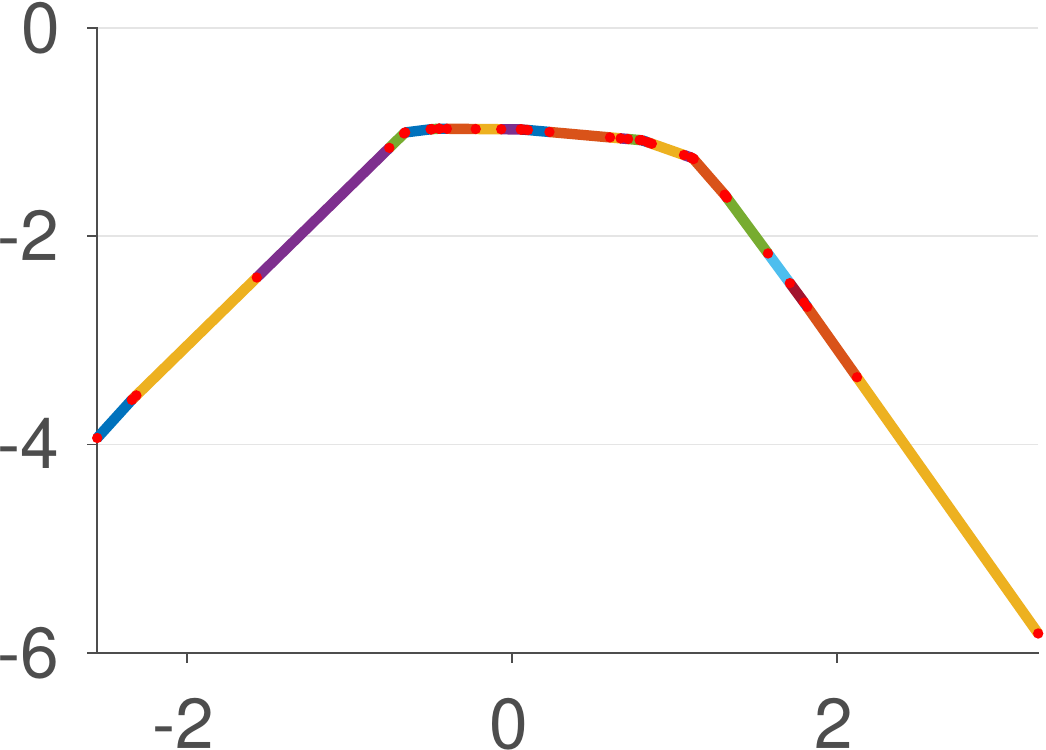}}
\hspace{0.5cm}
\subfloat[][\centering \footnotesize{Our approach~(\SI{0.21}{s}), \par $l(\hat{\theta}) = -683.38$, $N_{n,d}=8$}]{\includegraphics[width=0.30\textwidth]{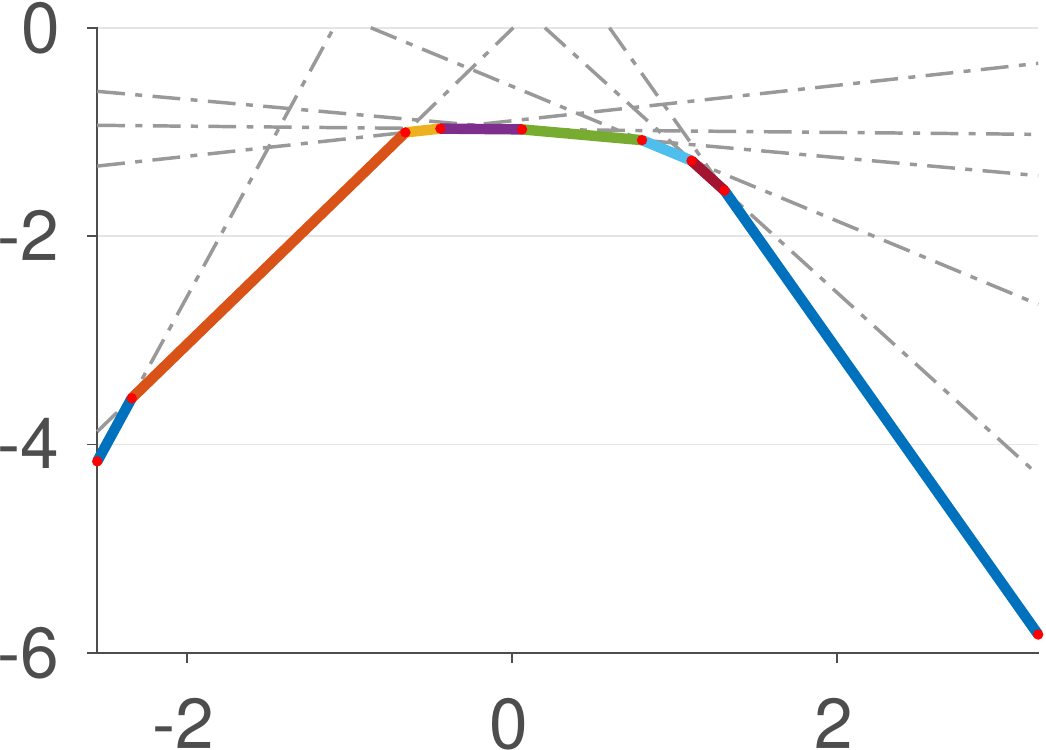}}

\subfloat[][\centering \footnotesize{D\"{u}mbgen et al.~(\SI{0.19}{s}), \par $l(\hat{\theta}) = -1096.30$, $N_{n,d}=6$}]
{\includegraphics[width=0.29\textwidth]{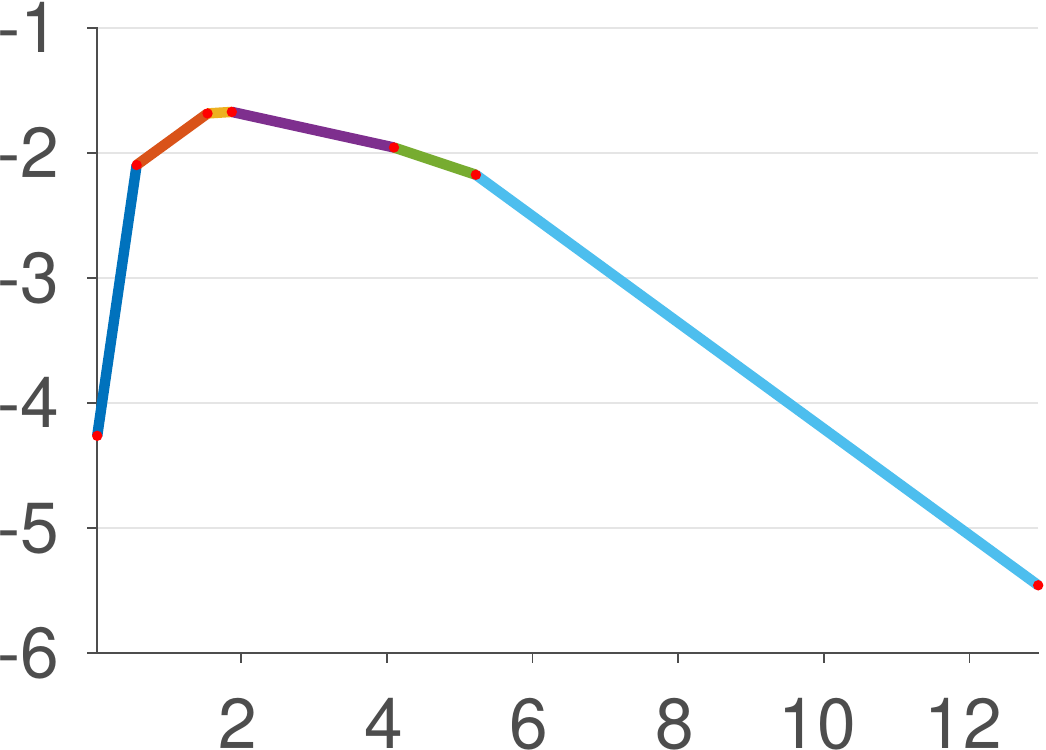}}
\hspace{0.5cm}
\subfloat[][\centering \footnotesize{Cule et al.~(\SI{3.54}{s}), \par $l(\hat{\theta}) = -1096.31$, $N_{n,d}=30$}]
{\includegraphics[width=0.29\textwidth]{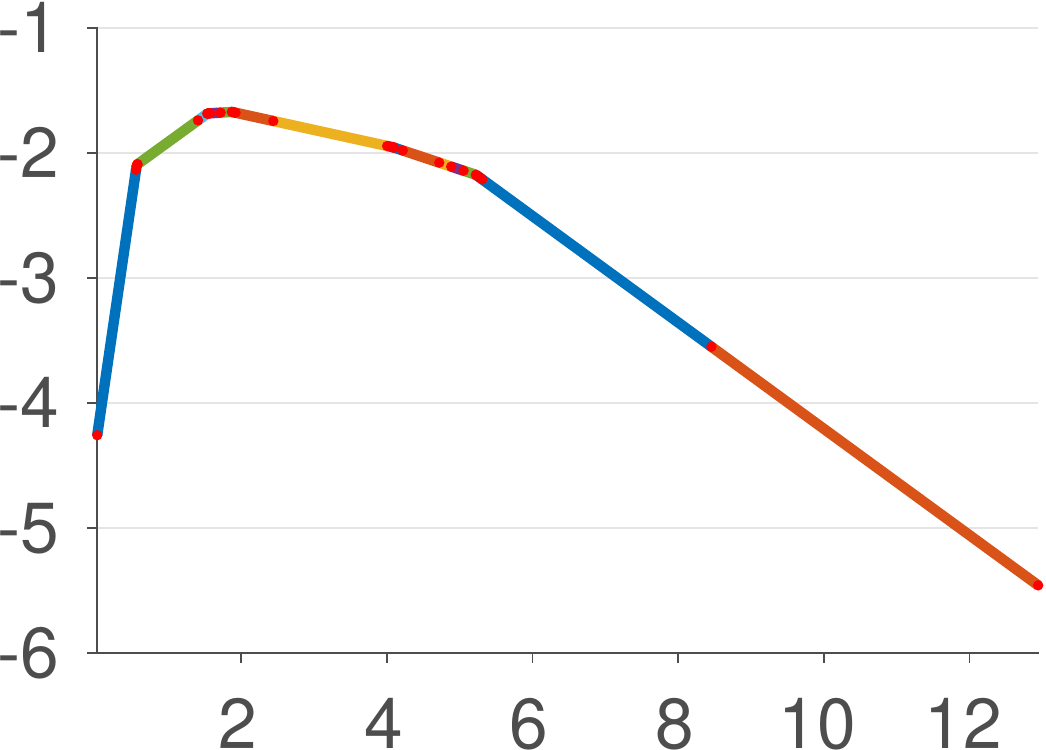}}
\hspace{0.5cm}
\subfloat[][\centering \footnotesize{Our approach~(\SI{0.27}{s}), \par $l(\hat{\theta}) = -1096.31$, $N_{n,d}=6$}]{\includegraphics[width=0.30\textwidth]{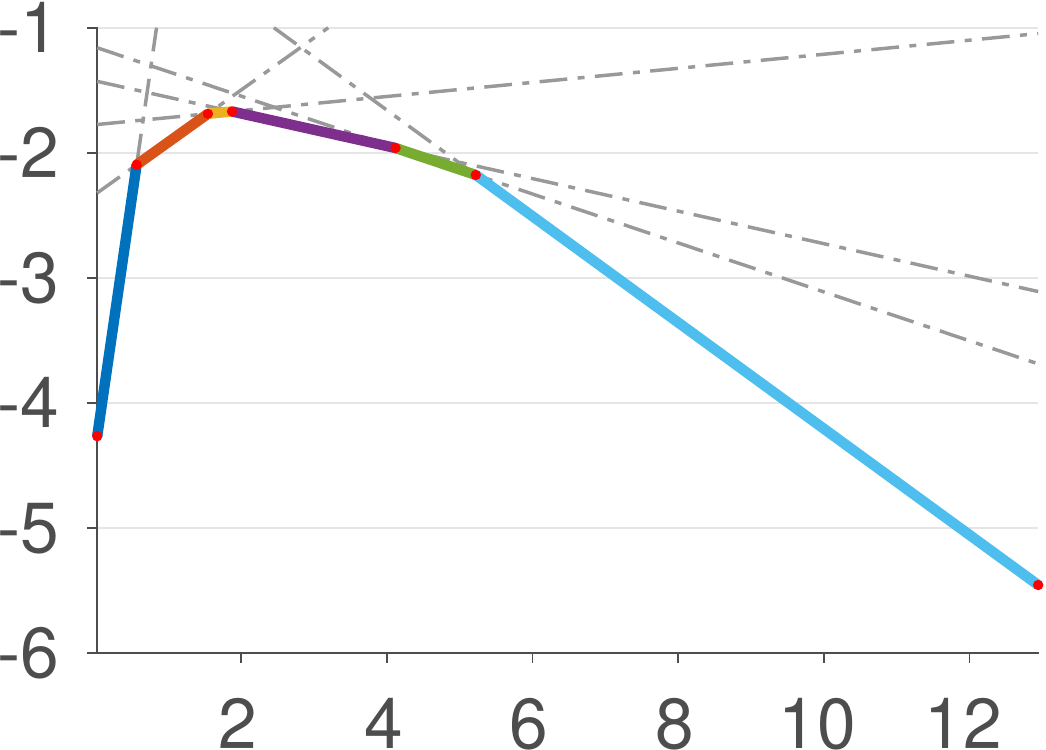}}

\subfloat[][\centering \footnotesize{D\"{u}mbgen et al.~(\SI{0.05}{s}), \par $l(\hat{\theta}) = -512.57$, $N_{n,d}=2$}]
{\includegraphics[width=0.29\textwidth]{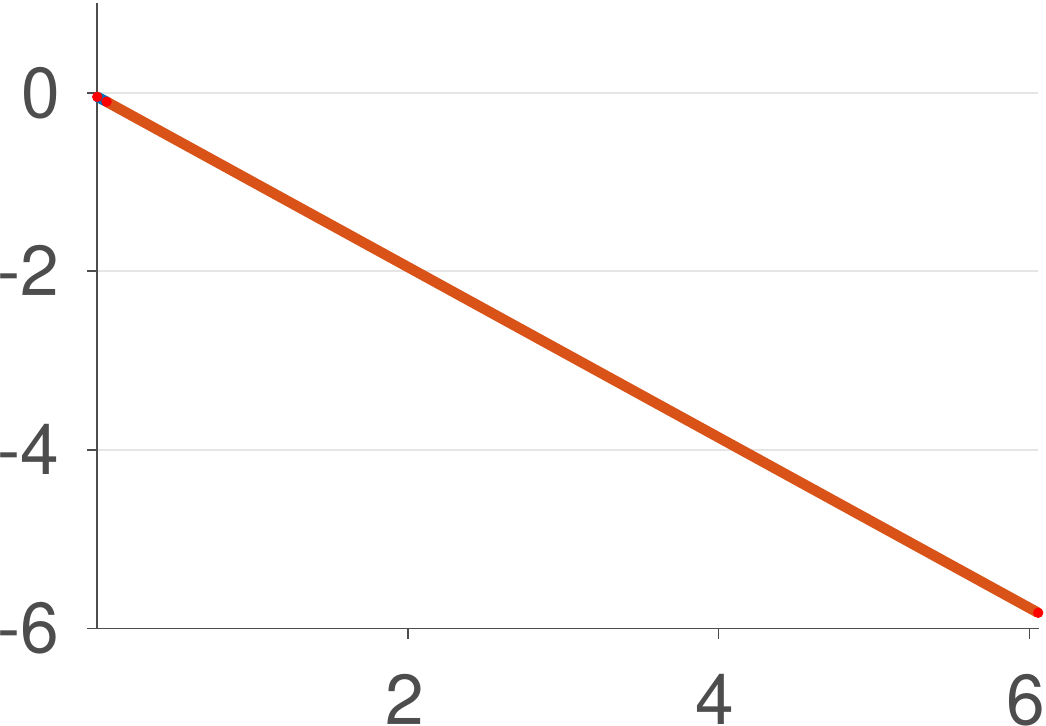}}
\hspace{0.5cm}
\subfloat[][\centering \footnotesize{Cule et al.~(\SI{3.79}{s}), \par $l(\hat{\theta}) = -512.59$, $N_{n,d}=8$}]
{\includegraphics[width=0.29\textwidth]{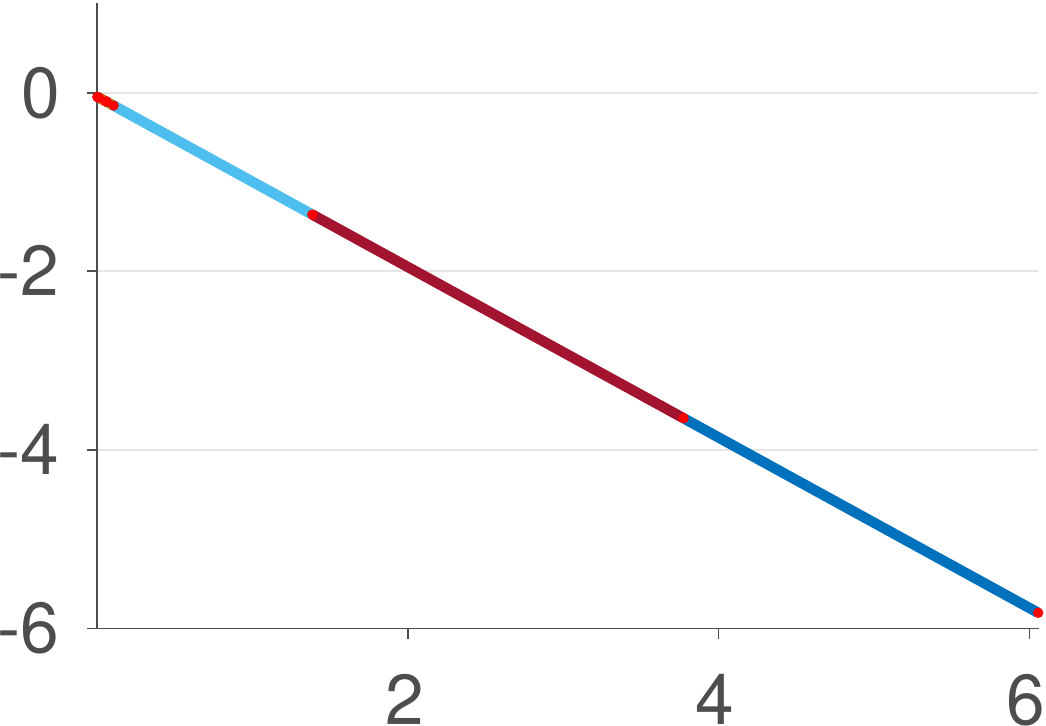}}
\hspace{0.5cm}
\subfloat[][\centering \footnotesize{Our approach~(\SI{0.09}{s}), \par $l(\hat{\theta}) = -512.57$, $N_{n,d}=1$}]{\includegraphics[width=0.30\textwidth]{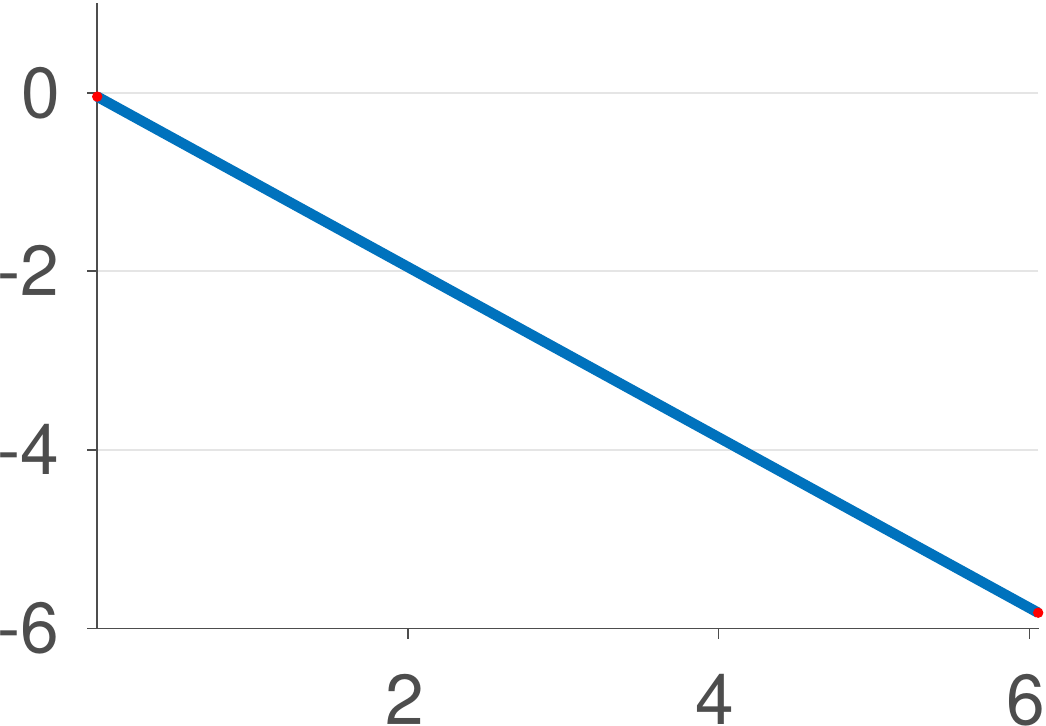}}
\caption{\new{Estimates $-\hat{\vphi}_n(x)$ and their piecewise linear representation for 500 samples drawn from a (a-c) normal distribution $\mc{N}(0,1)$, (d-f) gamma distribution $\Gamma(2,2)$ and (g-i) exponential distribution with $\lambda = 1$}. (a,d,g) The approach of \cite{duembgen2007} returns ground truth for univariate data, that is the maximal log-likelihood $l(\hat{\theta})$ and the correct piecewise linear representation of $-\hat\vphi_{n}$. (b,e,h) The approach of \cite{cule2010} also returns the optimal value $l(\hat{\theta})$ but generally works with a redundant piecewise linear representation that significantly increases runtime. (c,f,i) Our approach is as efficient as the former and only misses components of the piecewise linear representation that have very small support and hence a negligible impact on $l(\hat{\theta})$.
}
\label{fig:result-1d}
\end{figure}

In order to illustrate the sparse structure of solutions determined by our approach, we investigated examples in one and two dimensions. We compared the results with \cite{cule2010} whose approach (implemented in the R package \texttt{LogConcDEAD}) finds optimal solutions in terms of $L(\theta)$, but cannot take advantage of any sparsity of the solution, since its representation is based on the fixed  triangulation induced by $\mc{X}_n$ (recall Figure \ref{fig:n-gon}). For $d=1$, we additionally compared  to the approach of \cite{duembgen2007} using their R package $\texttt{logcondens}$, which can be only applied to univariate data, but finds optimal solutions in terms of $L(\theta)$ \textit{and} utilizes the minimal representation of the solution in terms of the number $N_{n,d}$ of hyperplanes. \new{We sampled 500 data points from three different distributions in 1-D (normal, gamma and exponential) and 500 samples from a normal distribution $\mc{N}(0,I_2)$ in 2-D.}

Figure~\ref{fig:result-1d} depicts the results of all three approaches for univariate data in terms of $-\hat{\varphi}_n(x) = \log \hat{f}_n(x)$. While all solutions are almost identical in terms of the estimated density, their parametrizations differ. The solution of \cite{duembgen2007} is guaranteed both to have the optimal sparse structure in terms of the number of hyperplanes and to yield the optimal value $L(\hat{\theta})$. Comparing their solution to ours, we see that they are almost identical, with only hyperplanes missing that have very small support and negligible impact on $L(\hat{\theta})$ \new{and $l(\hat{\theta})$ respectively}. The solution of \cite{cule2010}, on the other hand, is densely parametrized. The runtimes reflect these different parametrizations. 

\begin{figure}
\centerline{
\subfloat[][\centering \footnotesize{$\hat{f}_n(x)$, Cule et al.~(\SI{14.35}{s}), $l(\hat{\theta}) = -1159.19$}]{\includegraphics[width=0.42\textwidth]{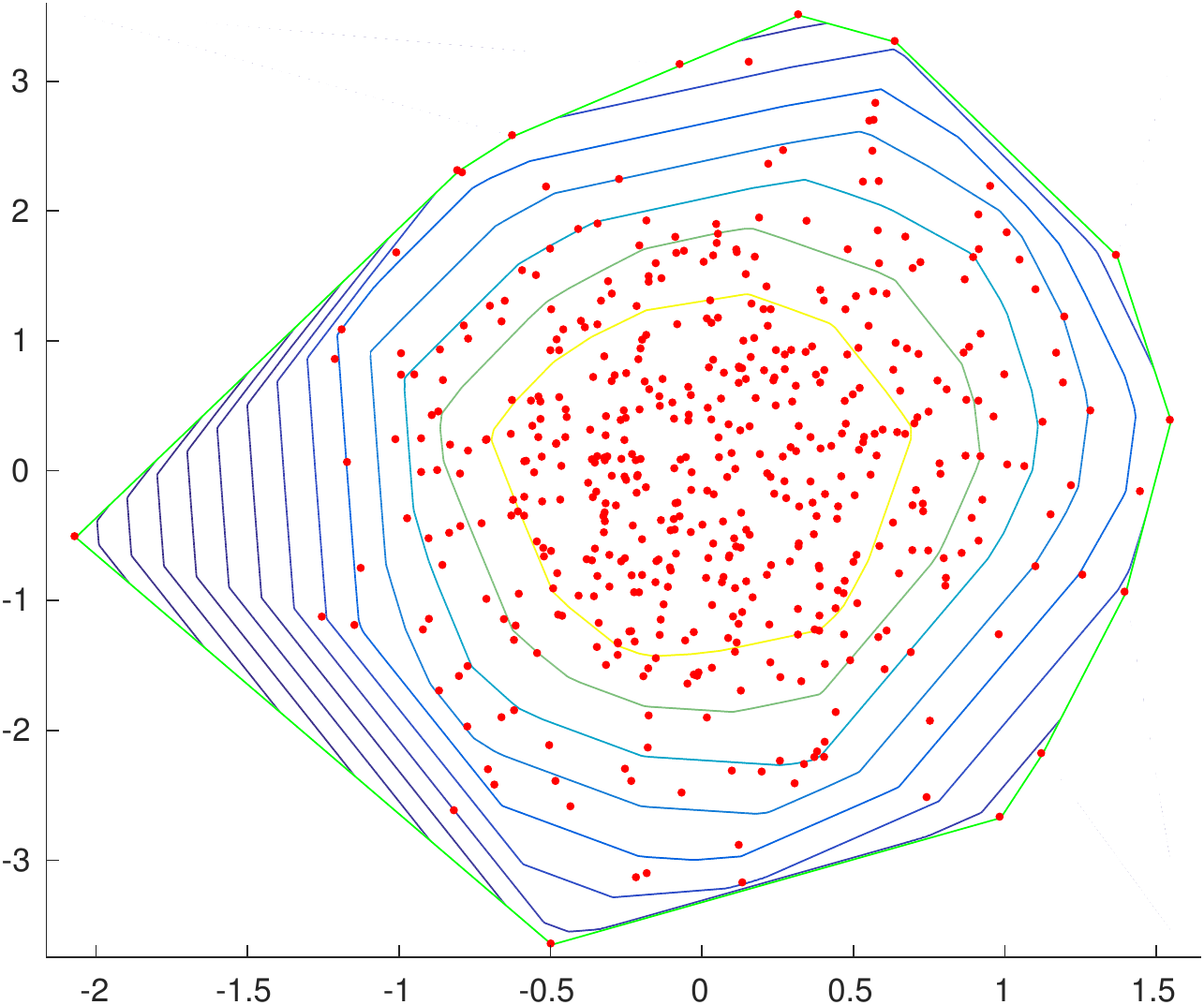}}
\hspace{2em}
\subfloat[][\centering \footnotesize{$\hat{f}_n(x)$, Our approach (\SI{0.16}{s}),  $l(\hat{\theta}) = -1159.53$}]{\includegraphics[width=0.42\textwidth]{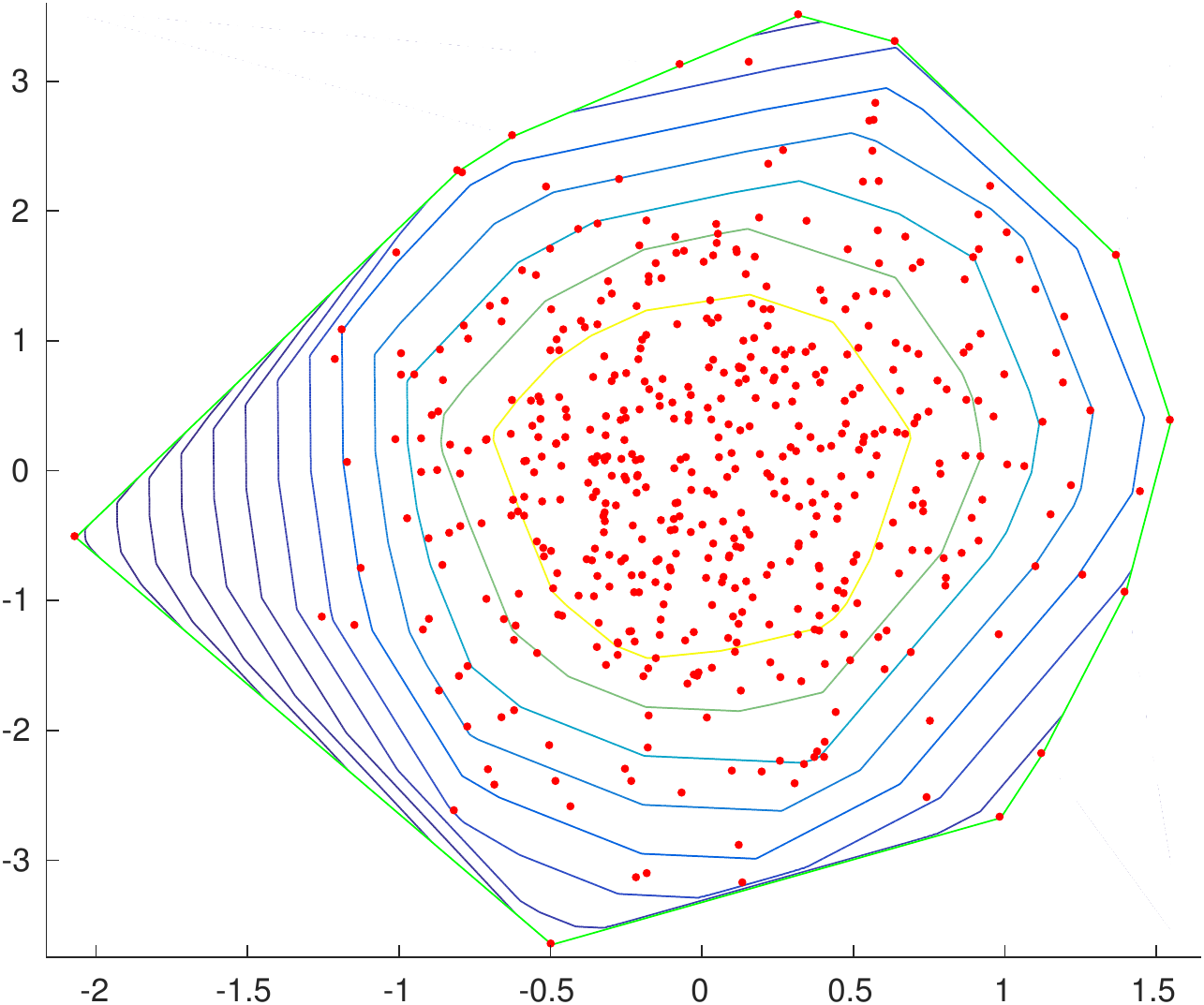}}
}
\vspace{0.3cm}
\centerline{
\subfloat[][$C_n, \;N_{n,d}=610$]{\includegraphics[width=0.42\textwidth]{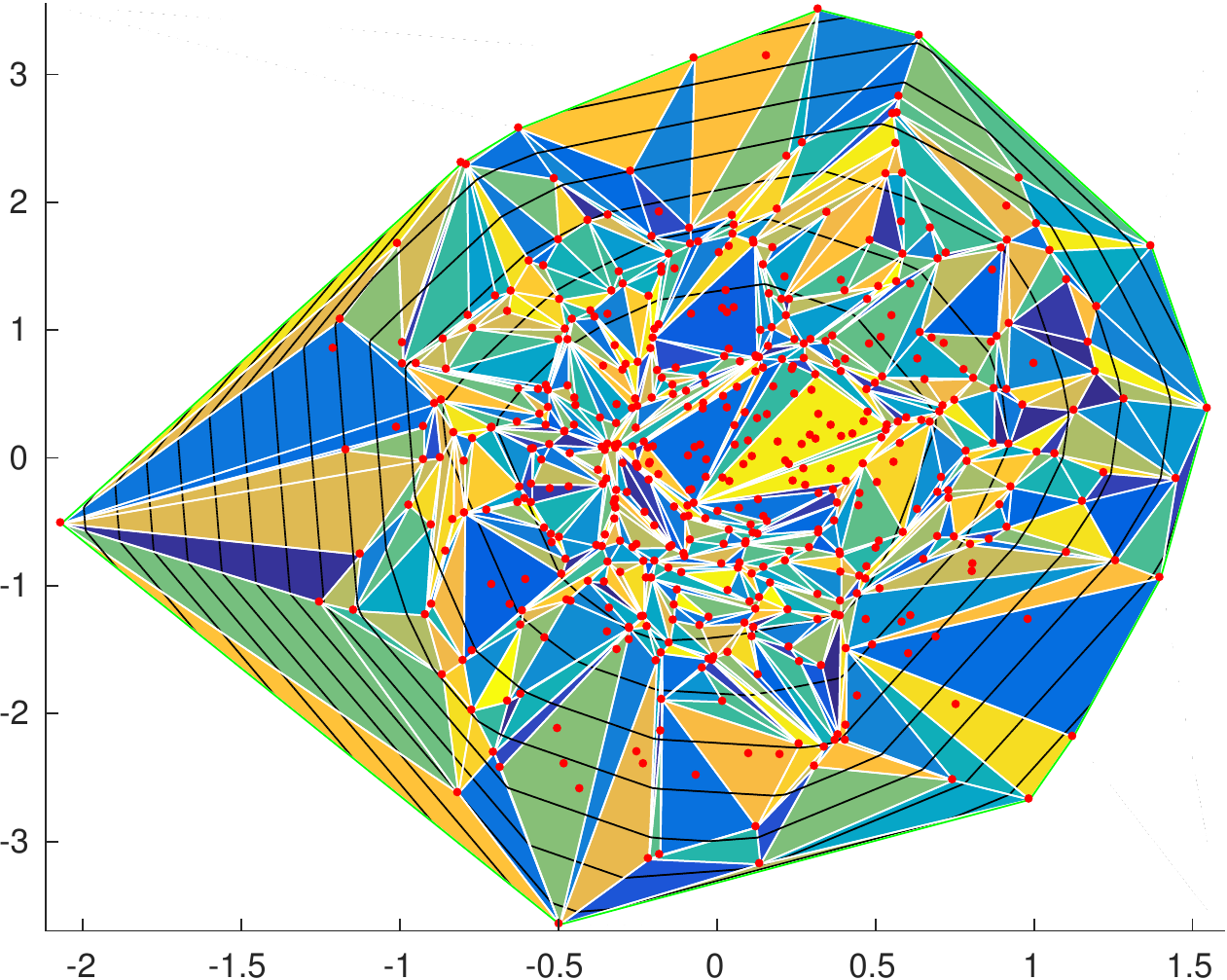}}
\hspace{2em}
\subfloat[][$C_n, \;N_{n,d}=66$]{\includegraphics[width=0.42\textwidth]{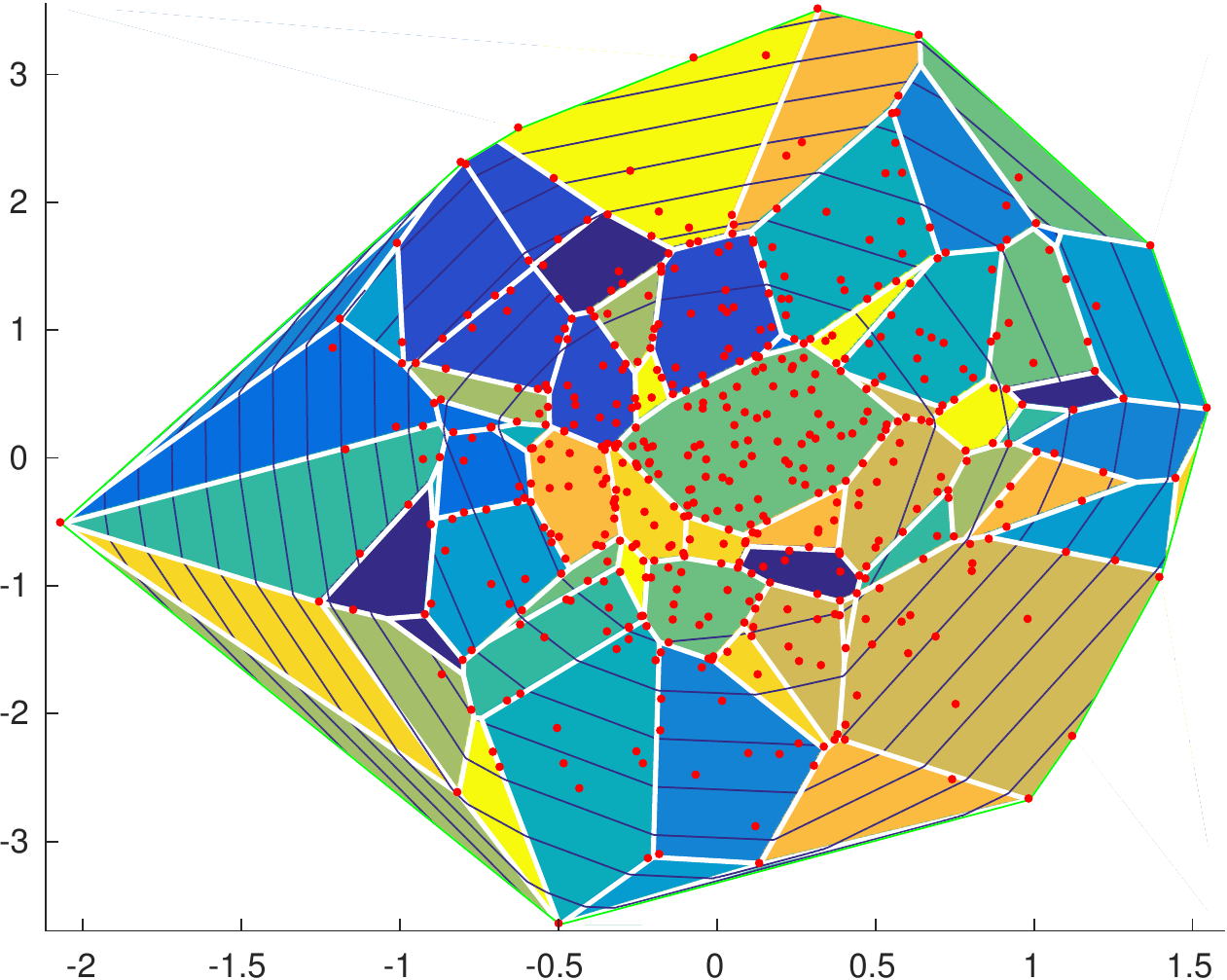}}
}
\caption{\textit{Top row}: Contour lines displaying estimates $\hat{f}_n(x) = \exp (\hat{\vphi}_n(x))$ for a sample of size $n=500$ drawn from $\mathcal{N}(0,I_2)$. The density plots (a) and (b) as well as the log-likelihood values $l(\hat{\theta})$ demonstrate that both solutions are very close to each other. \textit{Bottom row}: Decomposition of $C_n$ induced by the piecewise-linear concave functions $-\hat{\varphi}_n(x)$. While the approach of Cule et al.~induces a triangulation of $C_n$, our approach works with more general polytopes. Our approach adapts this representation to given data and thus avoids overfragmented representations as depicted by (c) that significantly increase runtime.
}
\label{fig:result-2d}
\end{figure}

We made similar observations for two-dimensional data $d=2$. Our approach found an density estimate $\hat{f}_n$ that is almost identical to the optimal solution but required only about $10\%$ of the parameters. Panels (a) and (b) of Figure \ref{fig:result-2d} depict the density $\hat{f}_n(x)$ estimated by the approach of \cite{cule2010} and our approach, respectively, whereas panels (c) and (d) show the respective decompositions of $\hat{\varphi}_n(x)$ into its affine representation $\hat{\vphi}_{i,n}$ \eqref{eq:vphi-affine}. While the solution of \cite{cule2010} is based on a fixed hyperplane arrangement and simplical supports $C_{n,i}$ induced by the given data, our decomposition of $C_n$ \eqref{eq:Cn-decomposition} is adaptive and can utilize more general convex polytopes $C_{n,i}$. Comparing panels (c) and (d) of Figure \ref{fig:result-2d} clearly shows the beneficial effect of adaptivity which is an `automatic' byproduct of minimizing $L(\theta)$ using our approach, together with pruning inactive hyperplanes.


\subsection{Quantitative Evaluation}
\label{sec:empirical-evaluation}
Besides the introductory experiments used for illustration, we conducted a comprehensive numerical evaluation using more challenging examples. The authors of \cite{cule2010} reported the evaluation of up to $n=2000$ data points and dimension $d=4$, drawn from $\mc{N}(0, I_d)$. In order to demonstrate the ability of our approach to process efficiently large data sets, we evaluated samples set of size up to $n=10000$ points and dimension $d=6$, drawn from the same distribution.

\begin{figure}
\centerline{
\subfloat[][Isotropic \texttt{O}]
{\includegraphics[width=0.26\textwidth]{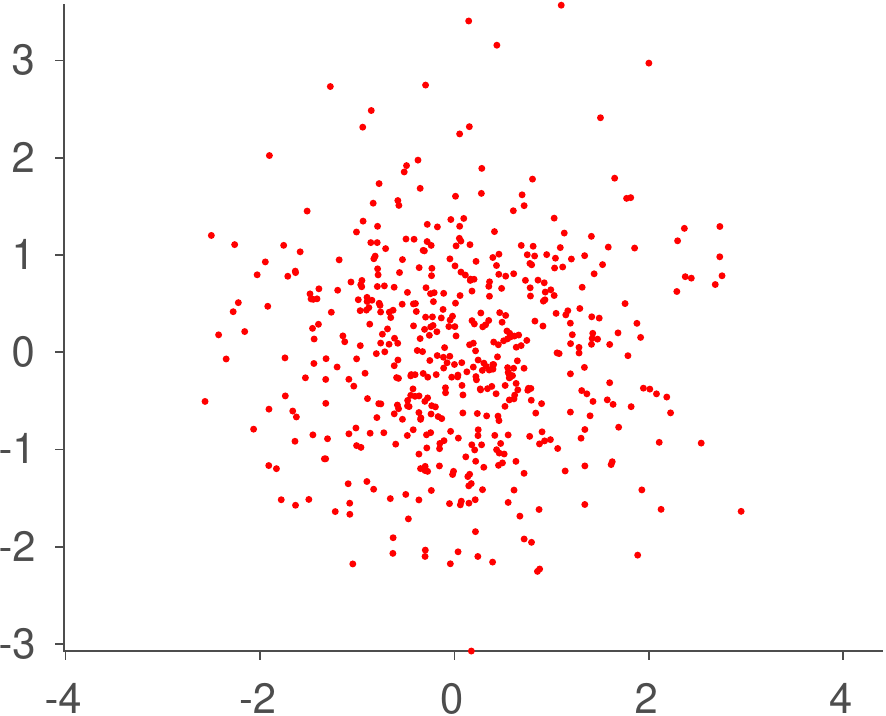}}
\hspace{2.5em}
\subfloat[][Anisotropic \texttt{I}]
{\includegraphics[width=0.26\textwidth]{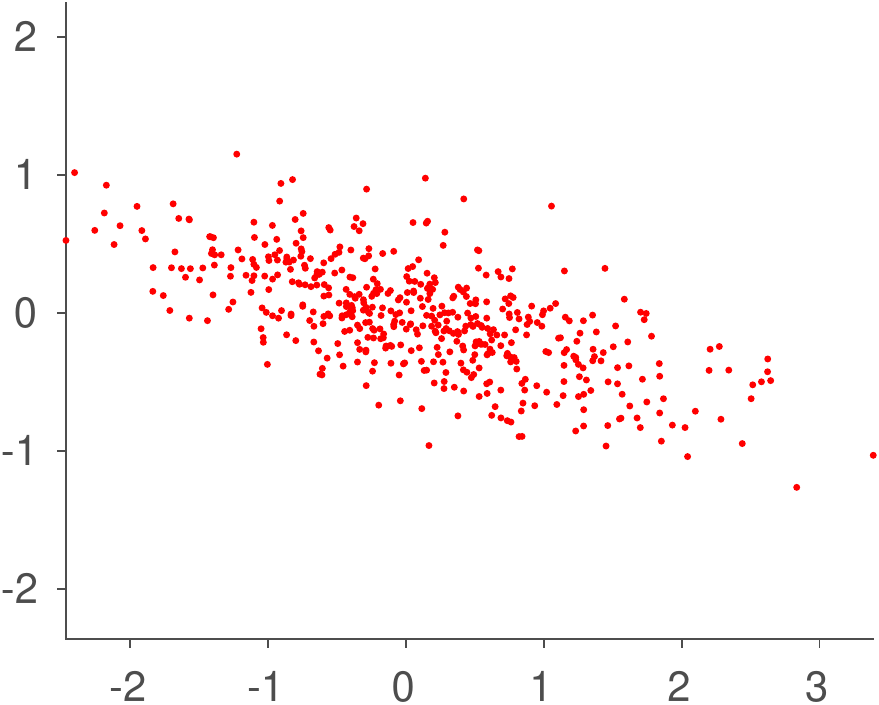}}
\hspace{2.5em}
\subfloat[][Anisotropic \texttt{II}]
{\includegraphics[width=0.26\textwidth]{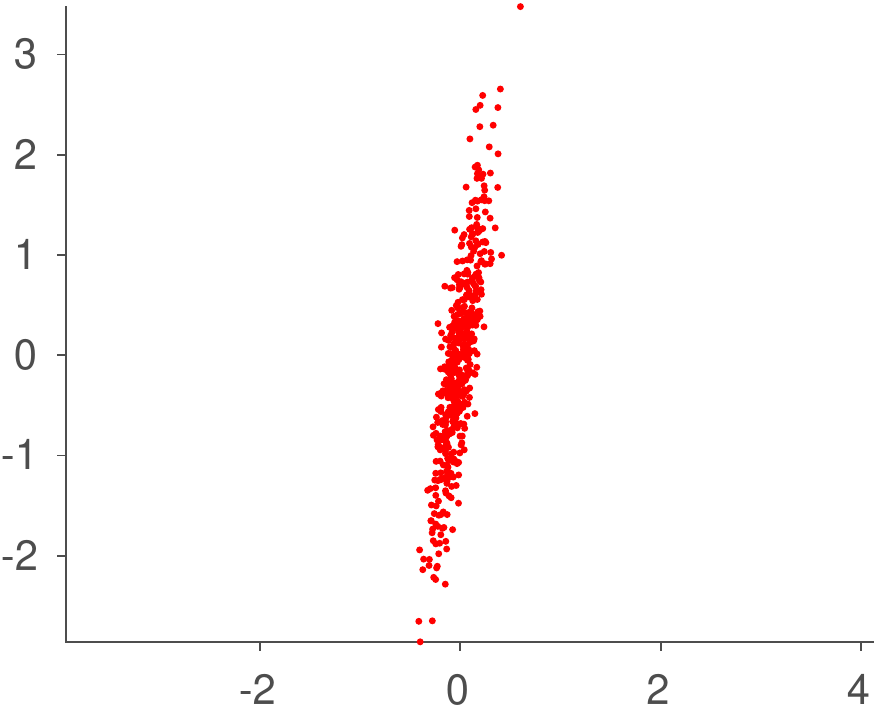}}
}
\caption{\newb{The three levels of anisotropy used in our experiments: None (a), mild (b) and strong (c).}}
\label{fig:anisotropic-levels}
\end{figure}
\begin{table}
\setlength{\tabcolsep}{0.1cm}
\renewcommand{\arraystretch}{1.05}
\centering
\caption{Runtimes for Cule's (mark: \texttt{C}) and our approach (\texttt{X}) and the resulting speedups. \new{For $d=1$ we additionally report results for the approach of D\"{u}mbgen et al.~(\texttt{D})}. \newb{Quality is the difference in total log-likelihood $l(\hat{\theta})$ to the solution of \texttt{C}. For the multidimensional datasets, accuracy is reported separately for all three levels of anisotropy (\texttt{O}, \texttt{I}, \texttt{II}). While accuracy slightly decreases for examples with strong anisotropy (degree \texttt{II}), it is still comparatively small given the number of samples.} For some samples we achieved a better log-likelihood value, because the implementation of Cule et al.~terminates after a hard-coded number of 15000 iterations which is no issue with our approach (for 2-D less than 500 iterations are required).}
\footnotesize
\vspace{4pt}
\begin{tabular} {c l c c c c c c c c}
\toprule
&  &  $n$ & 100 & 250 & 500 & \SI{1000}{} & \SI{2500}{} & \SI{5000}{} & \SI{10000}{}\\
\midrule
\multirow{6}{*}{1-D} & \multirow{3}{*}{Runtime} & \texttt{D} & \SI{0.08}{s}& \SI{0.2}{s}& \SI{0.3}{s}& \SI{0.4}{s}& \SI{0.7}{s}& \SI{2}{s}& \SI{3}{s}\\
&  & \texttt{X} & \SI{0.02}{s}& \SI{0.02}{s}& \SI{0.03}{s}& \SI{0.03}{s}& \SI{0.06}{s}& \SI{0.1}{s}& \SI{0.3}{s}\\
&  & \texttt{C} & \SI{0.2}{s}& \SI{0.7}{s}& \SI{3}{s}& \SI{25}{s}& \SI{16}{min}& \SI{1}{h} \SI{49}{min}& \SI{12}{h} \SI{52}{min}\\
& \textit{Speedup} && \textit{\SI{10}{x}}& \textit{\SI{43}{x}}& \textit{\SI{121}{x}}& \textit{\SI{780}{x}}& \textit{\SI{17695}{x}}& \textit{\SI{65513}{x}}& \textit{\SI{159534}{x}}\\
& \multirow{2}{*}{\textit{Quality}} & \texttt{D} & \textit{\SI{-0.0}}{}& \textit{\SI{-0.0}}{}& \textit{\SI{-0.0}}{}& \textit{\SI{-0.1}}{}& \textit{\SI{-0.1}}{}& \textit{\SI{-11.2}}{}& \textit{\SI{-69.1}}{}\\
&  & \texttt{X} & \textit{\SI{0.0}}{}& \textit{\SI{0.0}}{}& \textit{\SI{0.1}}{}& \textit{\SI{0.1}}{}& \textit{\SI{0.2}}{}& \textit{\SI{-10.4}}{}& \textit{\SI{-68.0}}{}\\
\midrule
\midrule
\multirow{4}{*}{2-D} & \multirow{2}{*}{Runtime} & \texttt{X} & \SI{0.5}{s}& \SI{0.5}{s}& \SI{0.6}{s}& \SI{0.7}{s}& \SI{0.8}{s}& \SI{1.0}{s}& \SI{2}{s}\\
&  & \texttt{C} & \SI{0.7}{s}& \SI{4}{s}& \SI{16}{s}& \SI{1}{min}& \SI{19}{min}& \SI{2}{h} \SI{17}{min}& \SI{14}{h} \SI{2}{min}\\
& \textit{Speedup} && \textit{\SI{1}{x}}& \textit{\SI{8}{x}}& \textit{\SI{26}{x}}& \textit{\SI{116}{x}}& \textit{\SI{1396}{x}}& \textit{\SI{8358}{x}}& \textit{\SI{27708}{x}}\\
& \multirow{3}{*}{\textit{Quality}} & \texttt{O}& \textit{\SI{0.1}}& \textit{\SI{0.2}}& \textit{\SI{0.3}}& \textit{\SI{0.7}}& \textit{\SI{1.0}}& \textit{\SI{1.3}}& \textit{\SI{-43.0}}\\
& & \texttt{I}& \textit{\SI{0.1}}& \textit{\SI{0.1}}& \textit{\SI{0.4}}& \textit{\SI{0.6}}& \textit{\SI{1.0}}& \textit{\SI{1.2}}& \textit{\SI{-44.9}}\\
& & \texttt{II}& \textit{\SI{0.1}}& \textit{\SI{0.3}}& \textit{\SI{0.8}}& \textit{\SI{1.3}}& \textit{\SI{1.4}}& \textit{\SI{3.3}}& \textit{\SI{0.5}}\\
\midrule
\multirow{4}{*}{3-D} & \multirow{2}{*}{Runtime} & \texttt{X} & \SI{3}{s}& \SI{6}{s}& \SI{6}{s}& \SI{7}{s}& \SI{8}{s}& \SI{10}{s}& \SI{13}{s}\\
&  & \texttt{C} & \SI{3}{s}& \SI{17}{s}& \SI{1}{min}& \SI{3}{min}& \SI{34}{min}& \SI{2}{h} \SI{58}{min}& \SI{18}{h} \SI{17}{min}\\
& \textit{Speedup} && \textit{\SI{1}{x}}& \textit{\SI{3}{x}}& \textit{\SI{11}{x}}& \textit{\SI{34}{x}}& \textit{\SI{248}{x}}& \textit{\SI{1028}{x}}& \textit{\SI{5209}{x}}\\
& \multirow{3}{*}{\textit{Quality}} & \texttt{O}& \textit{\SI{0.1}}& \textit{\SI{0.3}}& \textit{\SI{0.6}}& \textit{\SI{1.2}}& \textit{\SI{2.3}}& \textit{\SI{5.6}}& \textit{\SI{10.5}}\\
& & \texttt{I}& \textit{\SI{0.3}}& \textit{\SI{0.2}}& \textit{\SI{0.5}}& \textit{\SI{1.1}}& \textit{\SI{3.0}}& \textit{\SI{5.5}}& \textit{\SI{12.1}}\\
& & \texttt{II}& \textit{\SI{0.6}}& \textit{\SI{0.4}}& \textit{\SI{0.7}}& \textit{\SI{2.0}}& \textit{\SI{3.8}}& \textit{\SI{9.5}}& \textit{\SI{23.0}}\\
\midrule
\multirow{4}{*}{4-D} & \multirow{2}{*}{Runtime} & \texttt{X} & \SI{11}{s}& \SI{14}{s}& \SI{23}{s}& \SI{32}{s}& \SI{1}{min}& \SI{1}{min}& \SI{1}{min}\\
&  & \texttt{C} & \SI{13}{s}& \SI{1}{min}& \SI{5}{min}& \SI{24}{min}& \SI{2}{h} \SI{57}{min}& \SI{13}{h} \SI{16}{min}& -- \\
& \textit{Speedup} && \textit{\SI{1}{x}}& \textit{\SI{7}{x}}& \textit{\SI{16}{x}}& \textit{\SI{46}{x}}& \textit{\SI{200}{x}}& \textit{\SI{699}{x}}& --\\
& \multirow{3}{*}{\textit{Quality}} & \texttt{O}& \textit{\SI{0.5}}& \textit{\SI{0.7}}& \textit{\SI{0.5}}& \textit{\SI{1.0}}& \textit{\SI{2.2}}& \textit{\SI{7.6}}& --\\
& & \texttt{I}& \textit{\SI{0.2}}& \textit{\SI{0.9}}& \textit{\SI{0.6}}& \textit{\SI{0.8}}& \textit{\SI{2.0}}& \textit{\SI{6.8}}& --\\
& & \texttt{II}& \textit{\SI{0.5}}& \textit{\SI{0.7}}& \textit{\SI{0.6}}& \textit{\SI{3.1}}& \textit{\SI{2.5}}& \textit{\SI{7.4}}& --\\
\midrule
\multirow{4}{*}{5-D} & \multirow{2}{*}{Runtime} & \texttt{X} & \SI{14}{s}& \SI{24}{s}& \SI{44}{s}& \SI{1}{min}& \SI{2}{min}& \SI{3}{min}& \SI{4}{min}\\
&  & \texttt{C} & \SI{1}{min}& \SI{9}{min}& \SI{29}{min}& \SI{1}{h} \SI{23}{min}& \SI{13}{h} \SI{4}{min}& -- & -- \\
& \textit{Speedup} && \textit{\SI{5}{x}}& \textit{\SI{23}{x}}& \textit{\SI{40}{x}}& \textit{\SI{51}{x}}& \textit{\SI{319}{x}}& --& --\\
& \multirow{3}{*}{\textit{Quality}} & \texttt{O}& \textit{\SI{0.3}}& \textit{\SI{0.8}}& \textit{\SI{0.9}}& \textit{\SI{1.1}}& \textit{\SI{2.8}}& --& --\\
& & \texttt{I}& \textit{\SI{0.5}}& \textit{\SI{0.9}}& \textit{\SI{1.0}}& \textit{\SI{1.3}}& \textit{\SI{3.4}}& --& --\\
& & \texttt{II}& \textit{\SI{0.4}}& \textit{\SI{1.4}}& \textit{\SI{2.2}}& \textit{\SI{2.2}}& \textit{\SI{6.4}}& --& --\\
\midrule
\multirow{4}{*}{6-D} & \multirow{2}{*}{Runtime} & \texttt{X} & \SI{46}{s}& \SI{1}{min}& \SI{2}{min}& \SI{6}{min}& \SI{13}{min}& \SI{19}{min}& \SI{35}{min}\\
&  & \texttt{C} & \SI{9}{min}& \SI{1}{h} \SI{14}{min}& \SI{3}{h} \SI{6}{min}& \SI{10}{h} \SI{8}{min}& -- & -- & -- \\
& \textit{Speedup} && \textit{\SI{12}{x}}& \textit{\SI{41}{x}}& \textit{\SI{75}{x}}& \textit{\SI{93}{x}}& --& --& --\\
& \multirow{3}{*}{\textit{Quality}} & \texttt{O}& \textit{\SI{0.4}}& \textit{\SI{1.4}}& \textit{\SI{2.9}}& \textit{\SI{2.5}}& --& --& --\\
& & \texttt{I}& \textit{\SI{0.3}}& \textit{\SI{1.4}}& \textit{\SI{2.0}}& \textit{\SI{3.1}}& --& --& --\\
& & \texttt{II}& \textit{\SI{0.4}}& \textit{\SI{1.1}}& \textit{\SI{1.6}}& \textit{\SI{3.3}}& --& --& --\\
\bottomrule
\end{tabular}
\label{tab:results-all}
\end{table}

\newb{We extended their benchmarks in terms of sample size, dimension as well as introducing anisotropy to the covariance matrices: Besides the identity matrix (degree $\mathtt{0}$), we defined two levels of anisotropy ($\mathtt{I},\mathtt{II}$) based on the following metric for symmetric positive definitive matrices $S_1$ and $S_2$ \citep{Bhatia:2006aa}:
\begin{equation}
d(S_1,S_2) = \sqrt{\sum_{i=1}^d \Big(\log \lambda_i(S_1^{-1}S_2)\Big)^2}.
\end{equation}
Setting $S_1 = I_d$, we can see that the metric reduces to the euclidean norm of the logarithm of eigenvalues of $S_2$. We define our two levels of anisotropy as
\begin{equation}
\begin{aligned}
\mathtt{I}&: d(I_d, S_2) = 2.5,  \\
\mathtt{II}&: d(I_d, S_2) = 5.0.
\end{aligned}
\end{equation}  
Let $S_2$ be
\begin{equation}
S_2 = D E D^T, \qquad E = \diag(e),
\end{equation}
with $D$ being a \textit{random} orthogonal matrix. Since the eigenvalues of $S_2$ are solely determined by the vector $e$, we fix the last value of $e$ to 1 and set the remaining values to be equidistant in log-space, such that $d(I_d, S_2) = \| \log e \|_2 \in \{2.5, 5.0\}$. For dimension $d=2$ and $n = 500$, Figure~\ref{fig:anisotropic-levels} depicts examples for each degree of anisotropy.
}

For each level of anisotropy, we drew samples of sizes $n \in \{100, 250, 500, \allowbreak 1000, \allowbreak 2500, \allowbreak 5000, 10000\}$ in dimensions $d \in \{2,\ldots, 6\}$ and repeated each experiment five times. We run the R package \texttt{LogConcDEAD} (Version 1.6-1) by \cite{cule2010} in all dimensions but stopped increasing $n$ when the runtime exceeded 24 hours. Table \ref{tab:results-all} reports the results for all sample sizes, the speed ups as well as the quality of the solution achieved by our approach in comparison to the optimal solution returned by the approach of \cite{cule2010}, measured \newb{as the difference of total log-likelihoods $l(\hat{\theta})$}. 

\begin{figure}[t]
\centerline{
\subfloat
{\includegraphics[width=0.5\textwidth]{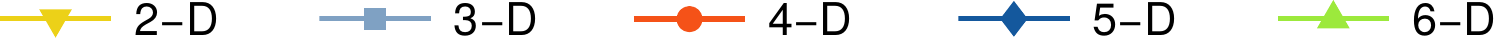}}
}
\setcounter{subfigure}{0}
\centerline{
\subfloat[][Cule et al.]
{\includegraphics[width=0.42\textwidth]{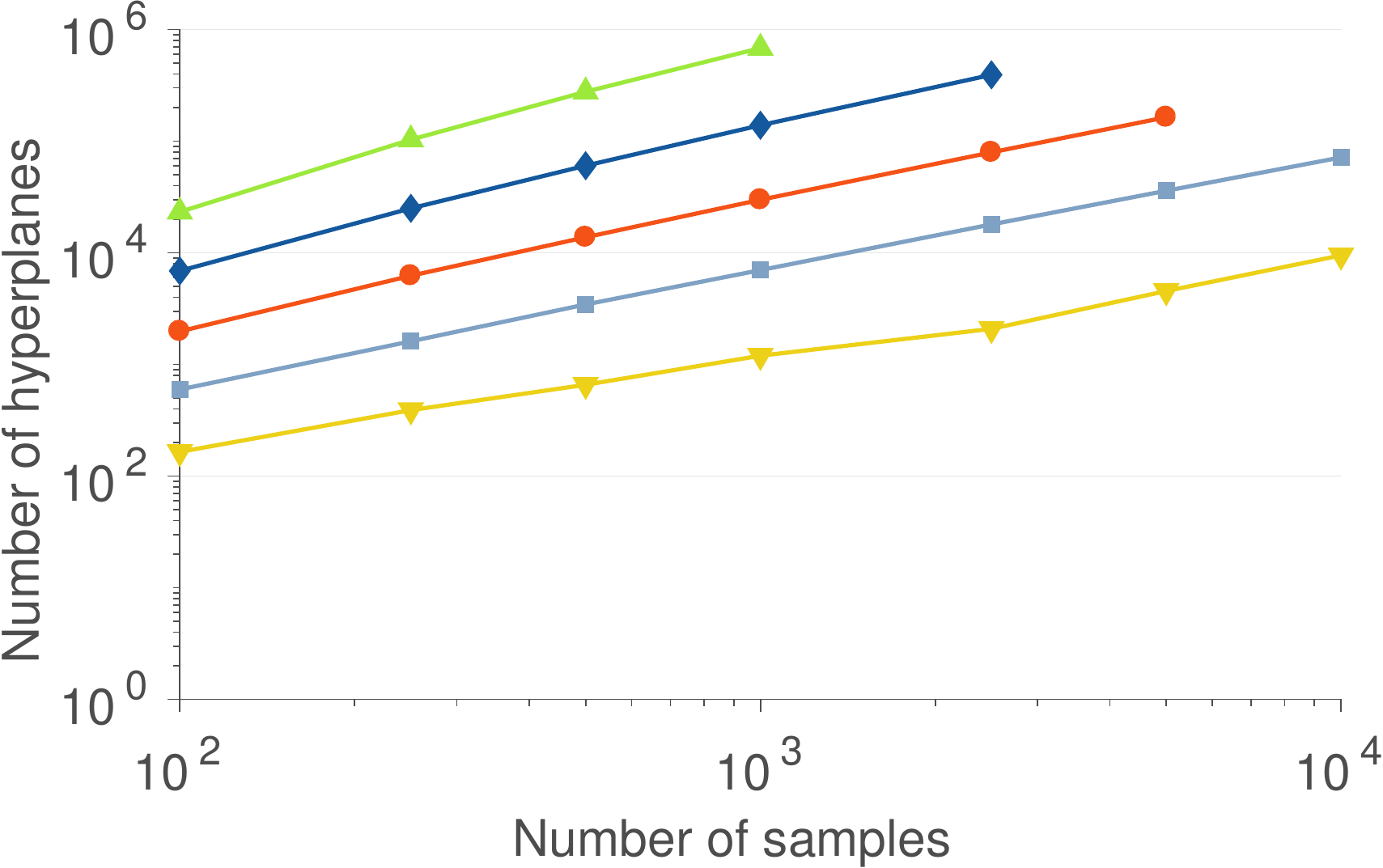}}
\hspace{3em}
\subfloat[][Our approach]
{\includegraphics[width=0.42\textwidth]{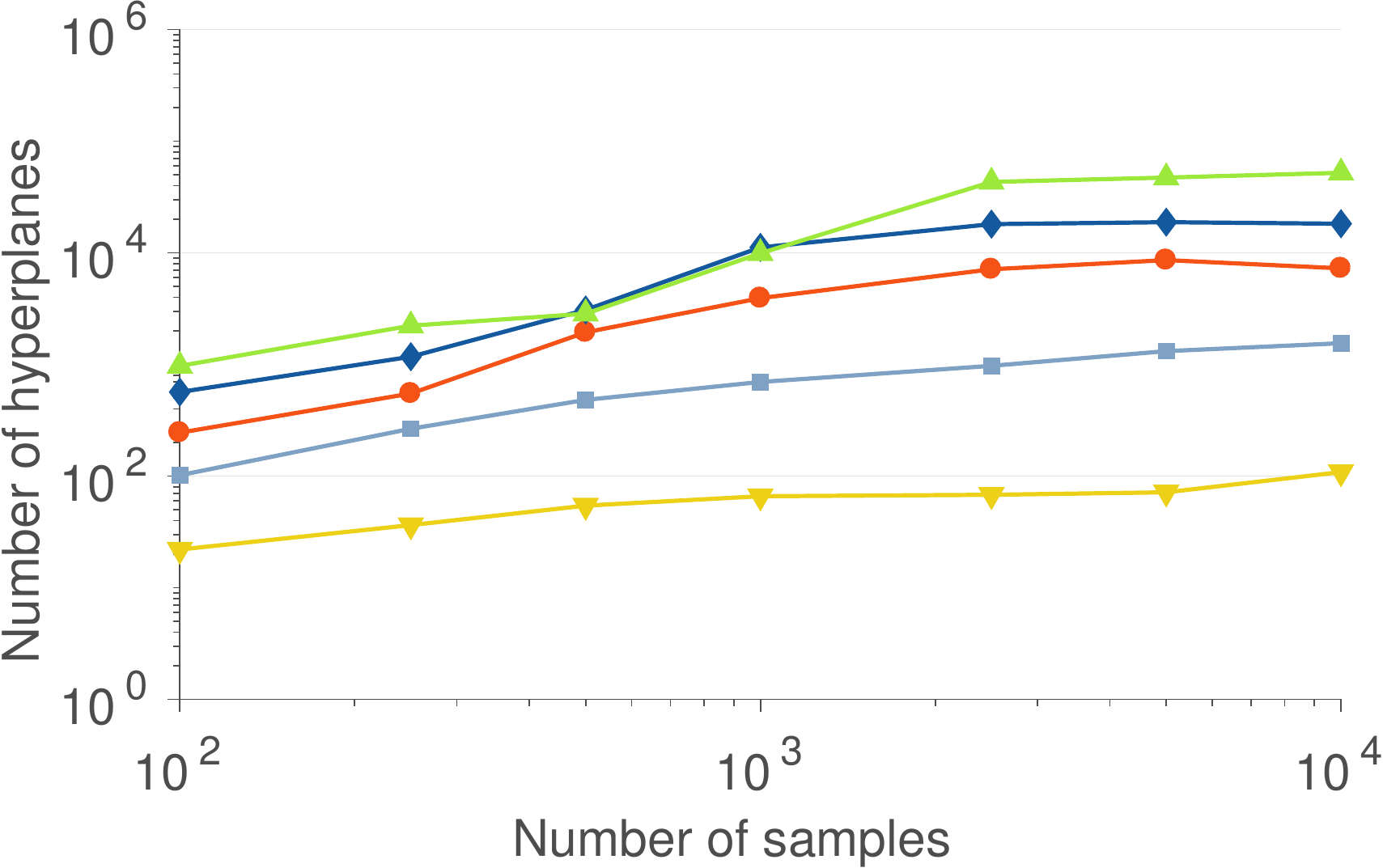}}}
\vspace{0.1cm}
\centerline{
\subfloat[][\texttt{X} (solid) \& \texttt{C} (dashed) vs.~true density]{\includegraphics[width=0.42\textwidth]{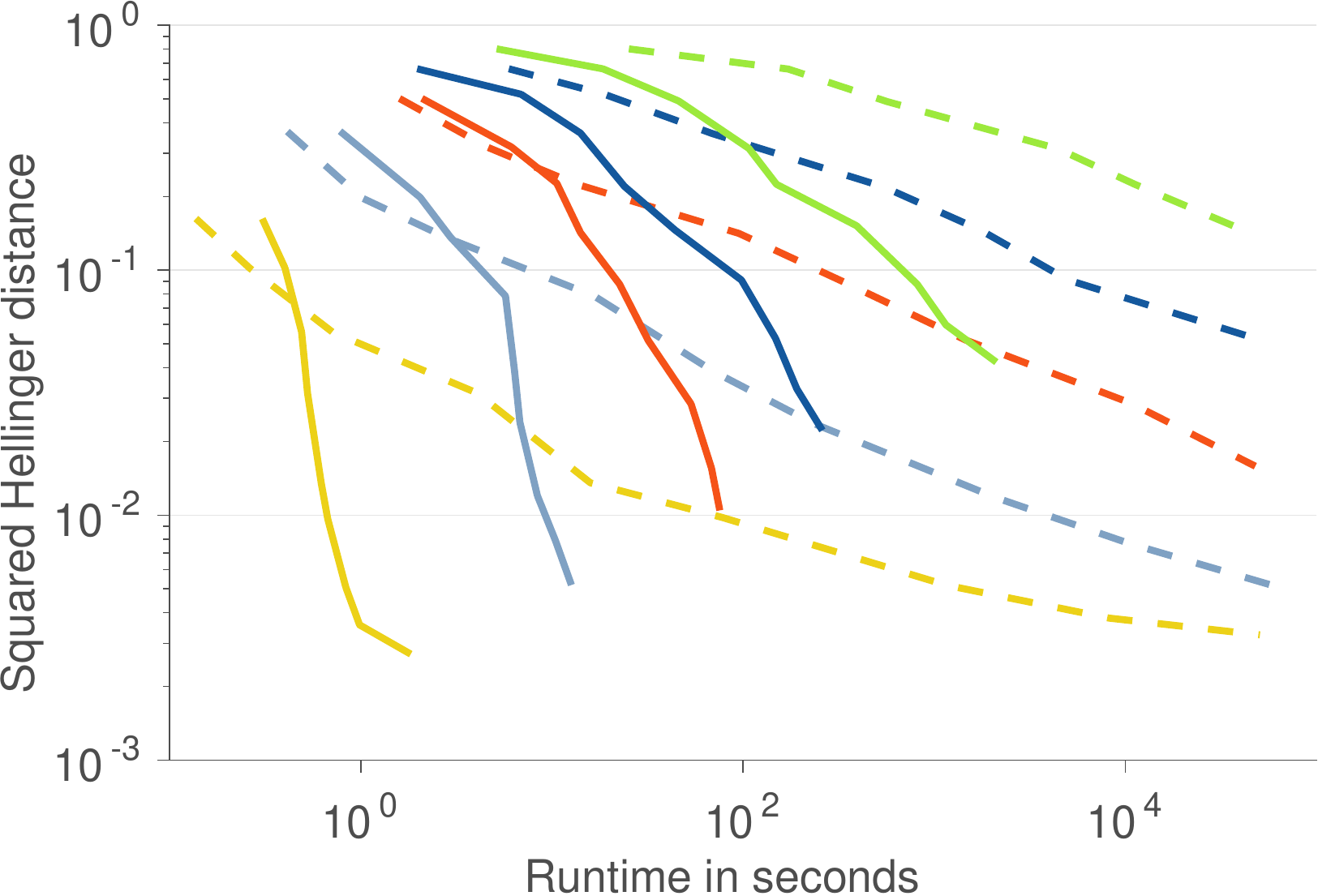}}
\hspace{3em}
\subfloat[][\texttt{X} vs.~\texttt{C} for degree \texttt{II} examples]{\includegraphics[width=0.42\textwidth]{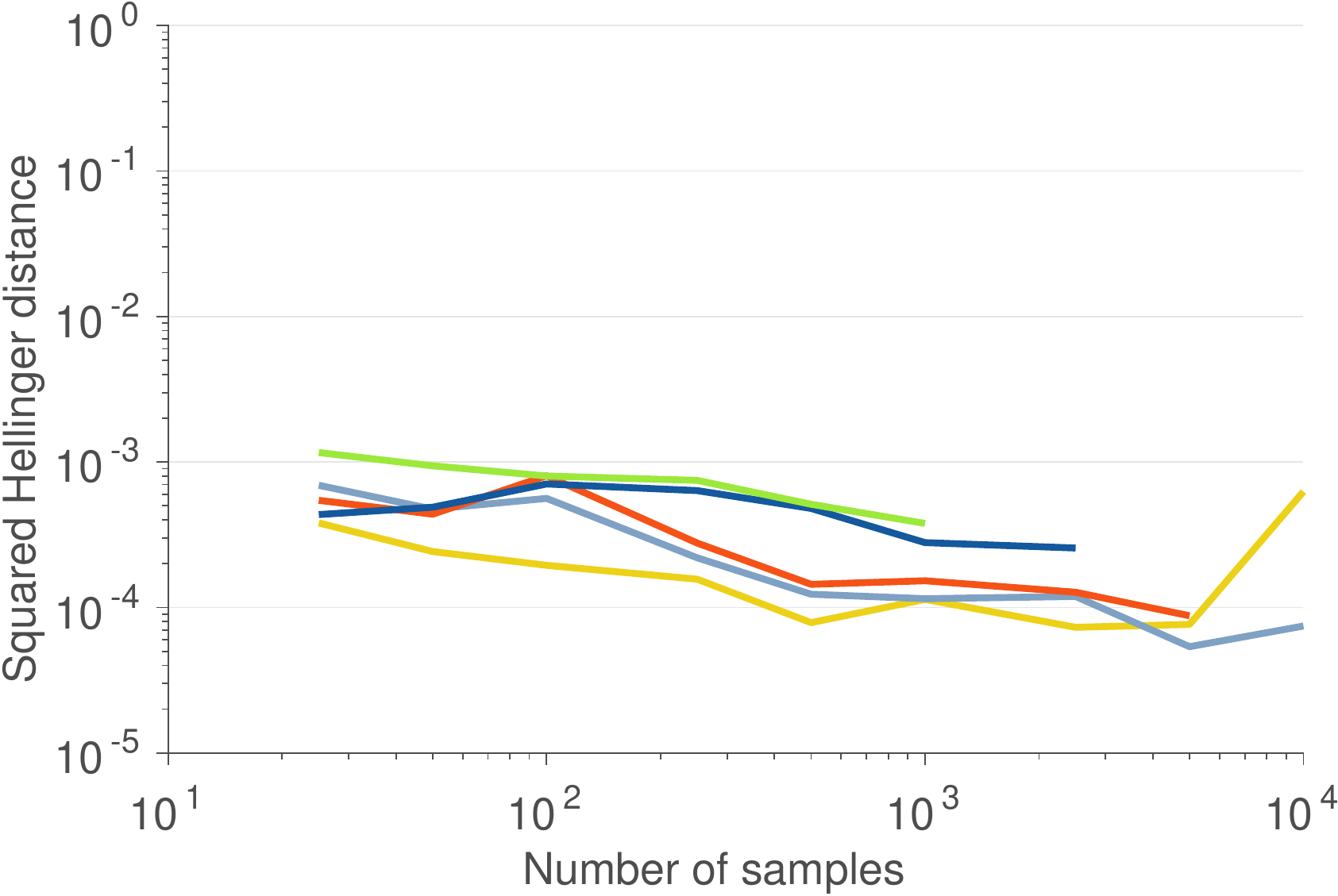}}
}
\caption{\textit{Top row}: Complexity of the density $\hat{f}_n$ in terms of the number $N_{n,d}$ of linear functions forming $-\hat{\varphi}_n(x) = \log\big(\hat{f}_n(x)\big)$. While scales linearly with the number of samples $n$ for the approach of Cule et al.~$N_{n,d}$, our approach shows only sublinear growth and starts to become flat for larger $n$. \textit{Bottom row}: Squared Hellinger  distance $h^2(\hat{f}_n, f)$ between (c) the estimate $\hat{f}_n$ and the underlying density $f$ versus runtime, evaluated for our test suite introduced in Section \ref{sec:empirical-evaluation} with additional results for $n=25$ and $n=50$. For a specific runtime, our approach (solid lines) obtains estimates closer to the underlying density $f$ while being able to process significantly more data points $n$. \newb{Panel (d) depicts the average squared Hellinger distance between our estimates and those of Cule et al.~for degree \texttt{II} samples (cf.~Figure \ref{fig:anisotropic-levels}). While constituting the most difficult samples, differences are very small and decrease with sample size.}}
\label{fig:numHypers}
\end{figure}

\newb{Overall we see that our approach delivers very accurate results, with very small differences given the respective number of samples. Regarding the influence of anisotropy on the accuracy, we notice only a slight increase the difference of $l(\hat{\theta)})$. Figure~\ref{fig:numHypers}~(d) provides a different perspective on the quality of the estimated densities, by showing the very small squared Hellinger distance between our estimates and those of \texttt{C} for degree \texttt{II} samples.}
While our approach estimated almost optimal densities, it achieved speed ups of up to a factor \SI{30000} (even more for $d=1$, though \cite{cule2010} point out that their approach is not designed with the one-dimensional case in mind). These factors increase with dimension and, in particular, with the number of data points.

We empirically observed and estimated how runtime $t$ depends on the sample size $n$. Regarding the approach \cite{cule2010} we observed linear dependency of the number of iterations on $n$, as is also mentioned in \cite{cule2010}. Moreover, we found that the complexity of their density estimate $\hat{f}_n$, expressed by the number of hyperplanes $N_{n,d}$, also grew linearly with $n$ (see Figure~\ref{fig:numHypers}~(a)), which is in accordance with our discussion in Section~\ref{sec:introduction}. Altogether runtime grew quadratically with the number of samples: $\mc{O}(n^2)$.

Examining the results for our approach revealed that the number of iterations depends much less on $n$. For example, for dimension $d=3$, the number of iterations for $n \in \{100, 1000, 100000\}$ was about the same, while for $d=4$ it doubled from $n=100$ to $n=1000$ but did not change when increasing $n$ to $10000$. This observation relates to the next paragraph below. Concerning the complexity of $\vphi_n(x)$ depending on the sample size $n$, we found that the number $N_{n,d}$ of hyperplanes grew sublinearly and started to become flat when $n > 1000$, c.f.~Figure~\ref{fig:numHypers}~(b).  All in all we found the average dependency of $t$ on $n$ to be roughly $O(\sqrt{n})$ for $d < 6$ and somewhat larger for $d=6$. 

The above observations partially relate to computational constraints for dimensions $d \geq 5$, where the grid size for numerical integration may limit the number $N_{n,d}$ of hyperplanes. For example, increasing the number of grid points by the factor 5 left unchanged the number of hyperplanes for $n=5000$ and $d=2$, increased it slightly by 20\% for $d=4$, but by about 50\% for $d=6$. Apparantly, this had no impact on the quality of the corresponding density estimates, but we cannot predict $N_{n,d}$ for $d > 4$. 



We conducted further experiments in order to demonstrate that our approach achieves both nearly optimal log-likelihoods and short runtimes: Based on the squared Hellinger distance
\begin{equation}
h^2(\hat{f}_n, f) = 1 - \int \sqrt{\hat{f}_n(x) f(x)} dx
\end{equation}
to measure the similarity between the estimate $\hat{f}_n$ and the true density $f$, Figure~\ref{fig:numHypers}~(c) depicts $h^2(\hat{f}_n, f)$ for the experiments performed in this section, plotted against the required runtime. Additional results for $n=25$ and $n=50$ are included. The plot demonstrates that our approach obtains much more accurate density estimates within a specific runtime. This  enables to take more data points into account than the approach of \cite{cule2010}, due to the adaptive sparse parametrization.

\subsection{Mixtures of Log-Concave Densities}\label{sec:Mixtures-Log-Concave} 
We extended our approach to the estimation of mixtures of log-concave densities. 
Let 
\begin{equation}
\Pi = \{\pi_1, \ldots, \pi_K\}, \qquad 
\sum_{k=1}^K \pi_k  = 1
\end{equation}
be the mixing coefficients for classes $1$ to $K$ and $\Theta = \{\theta_1, \ldots, \theta_K\}$ be class-specific hyperplane parameters. Then the log-concave mixture distribution is 
\begin{equation}
f_n(x|\Theta,\Pi) =  \sum_{k=1}^K \pi_k f_n (x |\theta_k),
\end{equation}
where $f_n(x|\theta_k)$ is a log-concave density as defined in Section \ref{sec:introduction}, parametrized by $\theta_k$. Given i.i.d samples $\mc{X}_n = \{x_1, \ldots, x_n\}$, maximizing the log-likelihood function
\begin{equation}\label{eq:L-mixture}
L(\Theta, \Pi) := \sum_{i=1}^{n} \log \sum_{k=1}^{K} \pi_{i} f_{n}(x_i| \theta_k) 
\end{equation}
is challenging due to the summation over $k$ inside the logarithm. A common technique to maximize locally $L(\Theta, \Pi)$ is the EM algorithm \citep{EM-dempster1977}. Here one introduces assignment probabilities 
\begin{equation}
\label{eq:responsibilities}
\gamma_{i,j} = \frac{\pi_j f_n(x_i|\theta_j)}{\sum_k \pi_k f_n(x_i|\theta_k)},\quad i = 1, \ldots, n,\quad
j = 1, \ldots, K
\end{equation}
that point $x_i$ belongs to class $j$
as latent parameters, that are iteratively estimated along with the mixture coefficients and the parameters of the mixture components. More specifically, the EM algorithm iterates the following two steps until convergence: The \textit{E-Step} computes \eqref{eq:responsibilities}. The \textit{M-Step} updates the mixing coefficients 
\begin{equation}
\pi_k = \frac{1}{n} \sum_i \gamma_{i,k}, \qquad k = 1, \ldots, K,
\end{equation}
and refines the parameters $\theta_k,\, k=1,\dotsc,K$ by minimizing the modified negative log-likelihood function
\begin{equation}
L_{\gamma}(\theta_k) := \frac{1}{n_k} \sum_{i=1}^n \gamma_{i,k} \vphi_{n, \gamma}(x_i) + \int_{C_n} \exp(-\vphi_{n, \gamma}(x)) dx, \quad n_k = \sum_i \gamma_{i,k},\quad k=1, \ldots, K.
\end{equation}
Since the objective function \eqref{eq:L-mixture} is non-convex, good initialization is essential. To obtain initial values for all $\gamma_{i,j}$, we follow \cite{cule2010} and use hierarchical clustering \citep{mclust-package}. We terminated the EM approach if the difference between $L(\Theta, \Pi)$ values in three subsequent iterations dropped below $10^{-5}$.

\begin{figure}
\centerline{
\subfloat[][USPS Dataset, $d=2$]
{\includegraphics[width=0.48\textwidth]{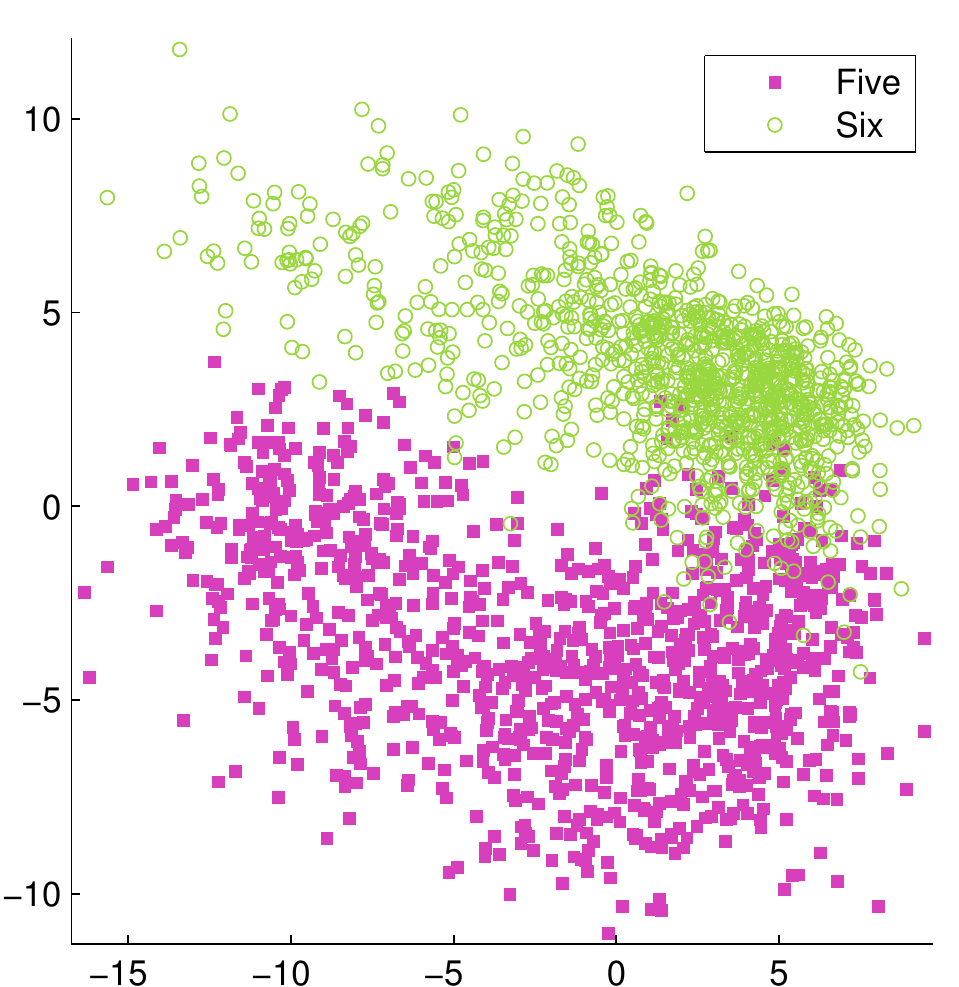}}
\hfill
\subfloat[][\centering GMM, $d=2$, 239~points~missclassified]
{\includegraphics[width=0.48\textwidth]{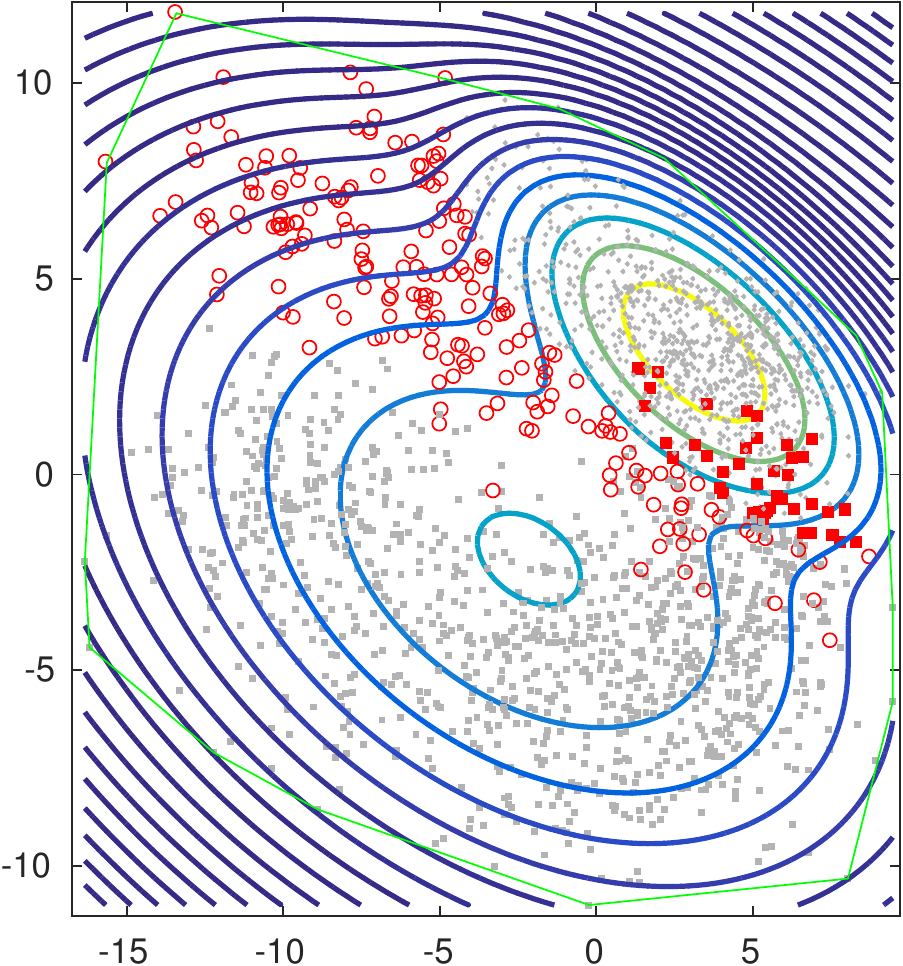}}}
\centerline{
\subfloat[][\centering Log-Concave, $d=2$, 200~points~missclassified]
{\includegraphics[width=0.48\textwidth]{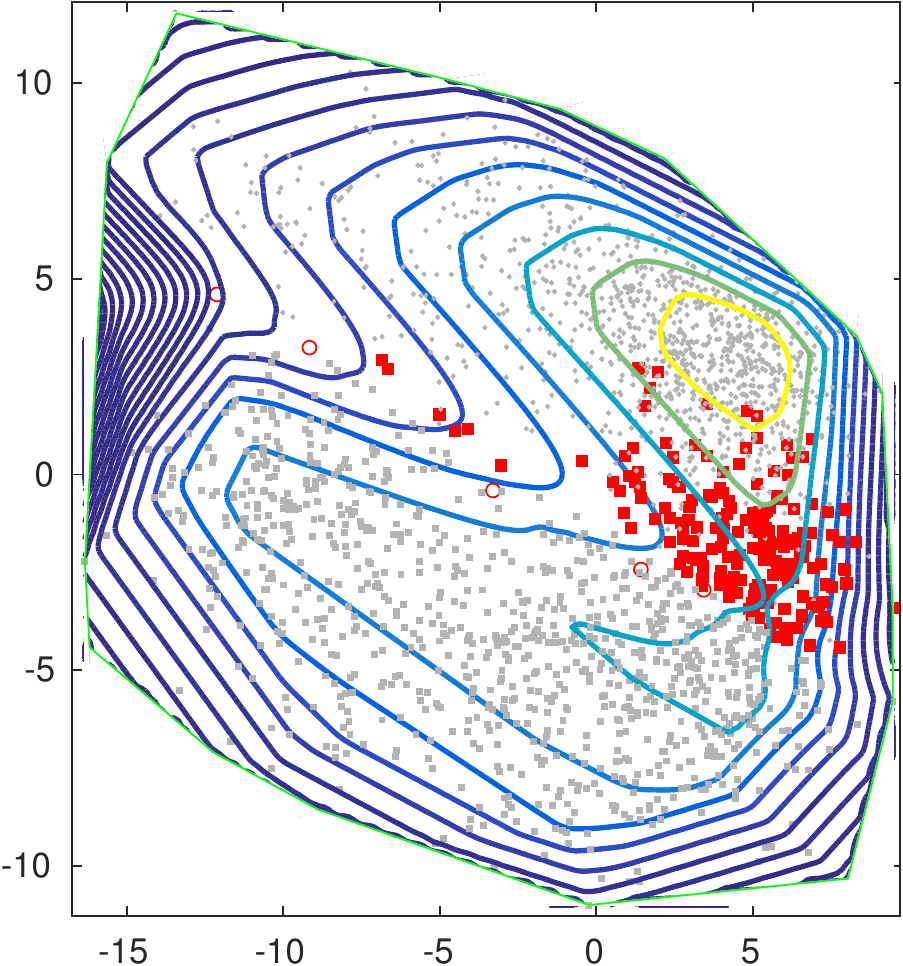}}
\hfill
\subfloat[][\centering Log-Concave, $d=3$, 78~points~missclassified]
{\includegraphics[width=0.48\textwidth]{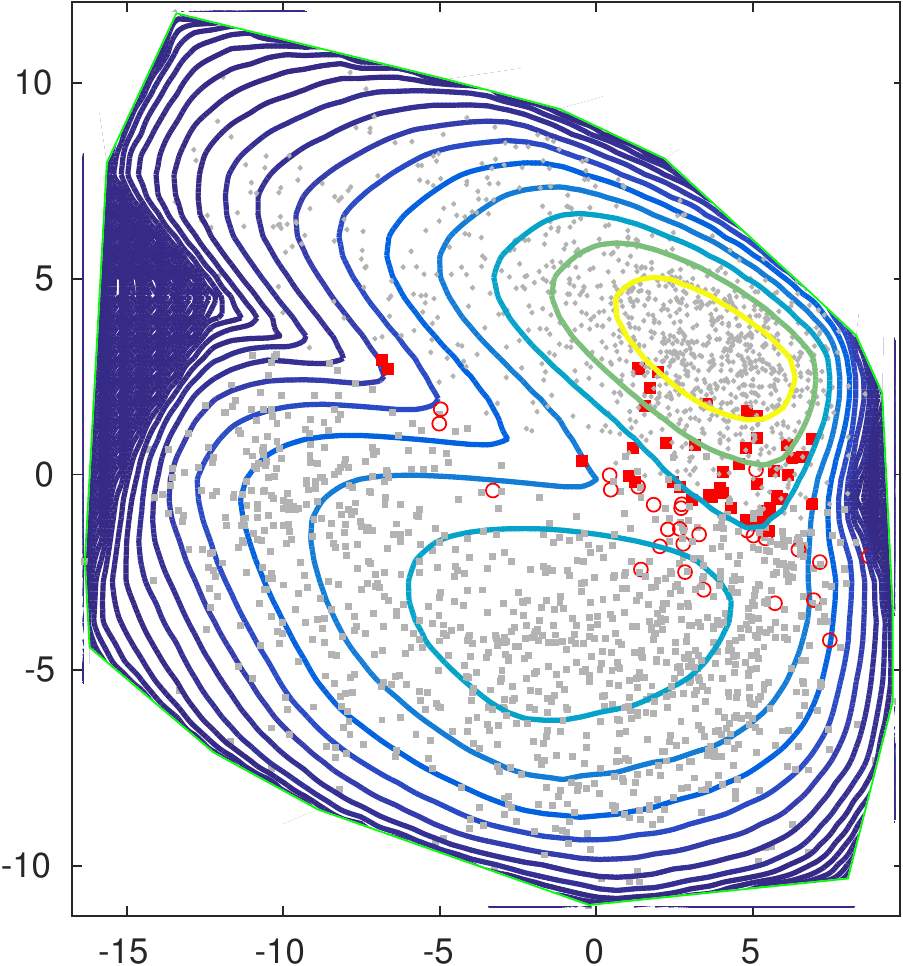}}}
\caption{(a) The USPS dataset projected onto the first two eigenmodes. (b,c) GMM and log-concave mixture estimates for this dataset. Missclassified points are drawn red. (d) Log-concave mixture estimate for $d=3$ with the third PCA dimension marginalized out by numerical integration for visualization.}
\label{fig:usps}
\end{figure}

We tested our approach on two datasets: 
\begin{itemize}
\item The \textbf{Wisconsin breast cancer dataset}, consisting of 569 samples with 30 features each, with 357 benign and 212 malignant instances.
\item The well-known \textbf{USPS dataset}, containing 11000 images of dimension $32 \times 32$ (i.e. 256 features) of handwritten digits from zero to nine. We selected all samples for the two classes `five` and `six`, 1100 each.
\end{itemize}
We reduced the dimension for both datasets to $d \in \{2,3\}$ using PCA. Figure \ref{fig:usps}~(a) depicts the USPS dataset projected onto the first two PCA eigenmodes. One can see the skewness of class 'six', which the Gaussian distribution is not able to capture. We compared our approach to \cite{cule2010} as well as to the standard Gaussian mixture model (GMM). Performance was measured in terms of the achieved log-likelihood and the number of misclassified samples, where each sample was assigned to the class $\mathrm{argmax}_k \pi_k f_n(x|\theta_k)$.

\begin{table}[t]
\setlength{\tabcolsep}{0.1cm}
\renewcommand{\arraystretch}{1.1}
\centering
\caption{Performance parameters for the estimation of log-concave mixture densities using the approach of \cite{cule2010} (mark: \texttt{C}) and our approach (mark: \texttt{X}). Estimates based on our approach are very close to the solutions of \texttt{C} in terms of the log-likelihood (\textit{Quality}) as well as the number of misclassified points, in one case even performing significantly better. Moreover, runtime is significantly reduced. Gaussian mixture models (\texttt{GMMs}) are clearly outperformed by log-concave mixtures for both datasets.}
\vspace{4pt}
\begin{tabular}{c r  r r  r r}
\toprule
& & \multicolumn{2}{c}{Wisconsin} & \multicolumn{2}{c}{USPS} \\ 
& & \multicolumn{1}{c}{2-D} & \multicolumn{1}{c}{3-D} & \multicolumn{1}{c}{2-D} & \multicolumn{1}{c}{3-D} \\
\midrule
\multirow{2}{*}{Runtime} &  \texttt{X} & \SI{21}{s}& \SI{3}{min}& \SI{38}{s}& \SI{20}{min}\\
& \texttt{C}& \SI{19}{min}& \SI{1}{h} \SI{3}{min}& \SI{2}{h} \SI{21}{min}& \SI{12}{h} \SI{50}{min}\\ 
\textit{Speedup} & & \SI{54}{x} & \SI{20}{x} & \SI{222}{x} & \SI{38}{x} \\
\textit{Quality} &  & \SI{2.6}& \SI{4.2}& \SI{1.5}& 15.3 \\
\multirow{2}{*}{Missclassified} & \texttt{X}& 47& 47&  \textbf{200}& \textbf{78}\\
& \texttt{C}& {\textbf{45}}& {\textbf{45}}& {201}& {109}\\
\midrule
\textit{Quality }& \texttt{GMM}& \SI{97.18}{\%}& \SI{93.36}{\%}& \SI{96.88}{\%}& \SI{93.93}{\%}\\
Missclassified & \texttt{GMM}& {58}& {64}& {239}& {215}\\
 \bottomrule
\end{tabular}
\label{tab:EM-all}
\end{table}

First, we compare Gaussian mixtures and log-concave mixtures. Table \ref{tab:EM-all} demonstrates that the log-concave mixture better reflects the structure of both datasets and clearly outperforms GMMs with respect to both performance measures. Naturally, using more information by increasing the dimension of the PCA subspace may lead to better estimates for the class-wise probabilities, as can be seen for the USPS dataset. Comparing the results of the log-concave mixture density estimates of both approaches, we again see very similar results in terms of the log-likelihood as well as the number of misclassified samples, the only exception being the USPS dataset for $d=3$. Again our approach achieves significant speedups. 

Figure~\ref{fig:usps}~(b)-(d) visualizes some mixture distributions for the USPS dataset for $d=2$ and $d=3$. While the GMM fails to properly model class `five` as expected, log-concave mixtures succeed, especially for $d=3$.

\section{Conclusion}
\label{sec:conclusion}
This work presented a new computational approach to the estimation of multivariate log-concave densities $\hat{f}_n$. Its crucial feature is the iterative search of a sparse representation of $\hat{f}_n$ in terms of the number of hyperplanes $N_{n,d}$ comprising the piecewise-linear concave function $\hat{\vphi}_n(x) = -\log \hat{f}_n (x)$. In addition, its parametrization involves hyperplanes supported on \textit{general} polytopes $C_{n,i}$, whereas the approach of \cite{cule2010} is restricted to simplices. By smoothing the original non-smooth objective, efficient established numerical methods can be applied. Overall, this led to significant speed ups compared to the state-of-the-art approach, in particular for larger data sets and increasing dimension. Although our approach minimizes a non-convex objective, we showed that this does not compromise the quality of the solution. On the contrary, almost optimal results are returned after runtimes that are shorter by factors up to \SI{30000}{x}.

Empirical evidence suggests the following dependency on the runtime $t$ and the sample size $n$: We observed that $t$ grows like  $\mc{O}(n^2)$ for the approach by \cite{cule2010}, resulting from a linear growth of both $N_{n,d}$ and the number of iterations with $n$. The observed linear dependency of $N_{n,d}$ on $n$ supports our reasoning in Section~\ref{sec:introduction} that the lower bound $\mc{O}(n)$ for $N_{n,d}$ (cf.~\cite{Dwyer:1991aa}) also applies to data sampled from log-concave distributions.

Regarding our approach we observed a $\mc{O}(\sqrt{n})$ dependency, due to a sublinear growth of $t$ with both $N_{n,d}$ and the number of iterations. We pointed out that, at least for $d \leq 4$, there is  strong empirical evidence that the sparse parametrization of $\hat{\vphi}_n(x)$ reflects structural properties of the maximum-likelihood estimator of log-concave densities, which the approach of \cite{cule2010} does \textit{not} exploit: Our approach successfully identifies maximal polytopes where  $\log\big(\hat{f}_n\big)$ is linear.
Since no theoretical results are available, to our knowledge, regarding the sparse parametrization of $\hat{\varphi}_n(x)$ for the case $d \geq 2$, our empirical results may stimulate corresponding research. 


We published our code with the \texttt{R} package \texttt{fmlogcondens} in the CRAN repository \citep{fmlogcondens}. A Matlab implementation is also available \\ at \url{https://github.com/FabianRathke/FastLogConvDens}.

\subsubsection*{Acknowledgements} This work has been supported by the German Research Foundation (DFG), grant GRK 1653, as part of the research training group on ``Probabilistic Graphical Models and Applications in Image Analysis'' \url{http://graphmod.iwr.uni-heidelberg.de}.
\appendix

\section{Calculating $\nabla L_{\gamma}(\theta^{(k)})$}
\label{sec:implementation}
A large amount of computation time is is spent on computing the gradient $\nabla L_{\gamma}(\theta^{(k)})$ of the objective \eqref{eq:def-L-theta-gamma}.
The gradient component with respect to a hyperplane normal $a_j$ reads
\begin{equation}
\nabla_{a_j} L_{\gamma}(\theta) = \frac{1}{n} \sum_{i=1}^n w_{ij} x_i - \Delta \sum_{l=1}^m w_{lj}  z_l \exp\big(-\vphi_{n,\gamma}(z_l)\big), \quad 
w_{ij} := \frac{\exp (\frac{1}{\gamma}(\la a_j, x_i \ra + b_j))}{\sum_{k=1}^{N_{n,d}} \exp(\frac{1}{\gamma}(\la a_k, x_i \ra  + b_k))}.
\end{equation}
Gradient components for the intercepts $b_j$ are similar.
Since $\frac{1}{\gamma}(\la a_j, x_i \ra  + b_j)) \to \infty$ for $\gamma \to 0$, a robust evaluation of terms $w_{ij}$ that prevents numerical overflow of the exp-function is given by
\begin{equation}
\begin{gathered}
w_{ij} = \frac{\exp (\nu_j)}{\sum_{k=1}^{N_{n,d}} \exp (\nu_k)} = \frac{\exp(\nu_j - \max \nu)}{ \sum_{k=1}^{N_{n,d}} \exp(\nu_k - \max \nu) },\\ 
\nu_k(x_i) := \gamma^{-1} (\la a_k, x_i \ra + b_k), \qquad
\nu = (\nu_1, \ldots, \nu_{N_{n,d}}).
\end{gathered}
\end{equation}
Similarly, the smooth max-approximation $\vphi_{n, \gamma}(x)$ \eqref{eq:vphi-smooth-approx} is numerically evaluated as
\begin{equation}
\vphi_{n,\gamma}(x) = \gamma \log \sum_{k=1}^{N_{n,d}} \exp(\nu_k) = \gamma \max \nu  + \gamma \log \sum_{k=1}^{N_{n,d}} \exp(\nu_k - \max \nu).
\end{equation}
Calculating $\nabla L_{\gamma}(\theta^{(k)})$ for all \new{combinations of} hyperplanes and grid points is the by far most expensive step in our approach. 
The problem is inherently sparse, however, since for most grid and data points only a few hyperplanes are relevant with most terms $w_{ij}$ negligibly small. 
We exploit this property in several ways. 

Computing the exponential function on a CPU is relatively expensive (about 50 times more than addition/multiplication \citep{ueberhuber2012numerical}). We set $\exp(\nu_j - \max \nu) = 0$, whenever $\nu_j - \max \nu \leq -25$. 
A second strategy attempts to reduce the number of hyperplane evaluations $\nu_k$. It utilizes two integration grids of varying density: A sparse grid to filter inactive hyperplanes and a dense grid to evaluate the integral of $f(x)$ and its gradient for all active hyperplanes. The sparse grid is divided into boxes $B_i$ consisting of $2^d$ adjacent grid points $\{v^i_1, \ldots, v^i_{2^d}\}$, e.g.~4 boxes in Figure \ref{fig:integration}~(a). For each box $B_i$ we perform the following steps:
\begin{enumerate}
\item Pick the point $z \in B_i$ that is closest to the mean of $\mc{X}_n$, evaluate all $\nu_k(z)$, $k=1, \ldots, N_{n,d}$ and set $\bar{k} = \argmax_k \nu_k$.
\item 
For each $k = 1, \ldots, N_{n,d}
$ find $\Delta_k$, the upper bound on the increase of hyperplane $k$ relative to hyperplane $\bar{k}$ in $B_i$.
Let $w_j$ be the width of box $B_i$ in dimension $j$ and 
\begin{equation}
\zeta_j = 
\begin{cases} 
      1 & z_j = \min_l v^i_{l,j}, \\
      -1 & \text{otherwise}.
   \end{cases}
\end{equation}
Then
\begin{equation}
\Delta_k  = \sum_{j=1}^d \max\left((a_{k,j} - a_{\bar{k},j})  w_j \zeta_j, 0\right).
\end{equation}
\item If $\nu_k(z) + \Delta_k - \nu_{\bar{k}}(z) \leq -25$, exclude hyperplane $k$ from the integration over $B_i$.
\end{enumerate}
For medium sized problems, this scheme reduces the number of evaluations of $w_{ij}$ by about 90\%.

Using a numerical integration scheme based on a regular grid facilitates \textit{parallelization}. We automatically distribute the numerical integration (and other expensive for-loops) across all available CPU cores, using the OpenMP API \citep{openmp-paper}. In addition, we utilize Advanved Vector Extensions (AVX), a technique that \textit{vectorizes} code by performing certain operations like addition or multiplication simultaneously for 8 floating point or 4 double values on a single CPU core. AVX is supported by all CPUs released since 2010. Both techniques, within-core and across-core parallelization led to speed ups by a factor of more than 10 on a standard four core machine. Due to the local character of most computations, transferring the code to the GPU promises even larger speed-ups.

\section{Exact integration}
\cite{cule2010} provided an explicit formula in order to evaluate exactly the integral in the objective \eqref{eq:def-J} based on a \textit{simplical decomposition} of $C_n$:
\begin{lemma}
Let $j = (j_0, \ldots, j_d)$ be a (d+1)-tupel of distinct indices in $\{1, \ldots, n\}$, such that $C_{n,i} = \conv(x_{j_0}, \ldots, x_{j_d})$ is a simplex with associated affine function $\vphi_{i,n} = \la a_i, x \ra + b_i$ and values $y_{j_l} = \vphi_{i,n}(x_{j_l})$ at its vertices. 
Then 
\begin{equation}
\label{eq:exact-integral}
\int_{C_{n,i}} \exp(-\vphi_{i,n}(x)) \dd{x} = \mathrm{vol}(C_{n,i}) \sum_{l=0}^d \exp(-y_{j_l}) \prod_{i \in \{0, \ldots, d \} \setminus{\{l\}}} (y_{j_i} - y_{j_l})^{-1}. 
\end{equation}
\label{lemma:integral-cule}
\end{lemma}
We apply this Lemma in Section \ref{sec:normalization} in order to normalize exactly the numerically computed density estimate $\hat{f}_n$ returned by our approach.

\bibliographystyle{elsarticle-harv}

\end{document}